\pdfoutput=1
\documentclass[12pt, preprint]{aastex}
\usepackage{fancyhdr}
\usepackage{longtable}
\usepackage{latexsym}
\usepackage{graphicx}
\usepackage{amsmath}
\usepackage{natbib}
\usepackage{upgreek}
\usepackage{multirow}
\usepackage{array}
\usepackage{pdflscape}
\usepackage{soul}

\newcommand{\PreserveBackslash}[1]{\let\temp=\\#1\let\\=\temp}
\newcolumntype{C}[1]{>{\PreserveBackslash\centering}p{#1}}
\newcolumntype{R}[1]{>{\PreserveBackslash\raggedleft}p{#1}}
\newcolumntype{L}[1]{>{\PreserveBackslash\raggedright}p{#1}}

\newcommand       \km           {\,{\rm km}}
\newcommand       \s            {\,{\rm s}}
\newcommand       \K            {\,{\rm K}}
\newcommand       \yr           {\,{\rm yr}}
\newcommand       \ppm          {\,{\rm ppm}}
\newcommand       \mum          {\,{\rm \mu m}}
\newcommand       \Ks           {{K_{\rm S}}}
\newcommand       \simali       {\,{\sim}}

\newcommand \taupd  {\tau_{\rm pd}}
\newcommand \taumc  {\tau_{\rm MC}}
\newcommand       \Teffqk      {\left(\frac{{T_{\rm eff}}}{1000{\rm K}}\right)}
\newcommand       \RV           {{R_V}}
\newcommand       \AV           {{A_{\rm V}}}

\newcommand       \AKs          {{A_{\rm{K_S}}}}

\newcommand       \gtsim        {\gtrsim}

\newcommand       \CJKs         {C_{\rm{JK_S}}}
\newcommand       \Teff         {T_{\rm {eff}}}

\shorttitle{Mid-Infrared Extinction Law}
\shortauthors{Xue et al.\ }

\begin{document}

\title{A Precise Determination of the Mid-Infrared Interstellar
       Extinction Law Based on the APOGEE Spectroscopic Survey}


\author{Mengyao Xue\altaffilmark{1},
B.~W. Jiang\altaffilmark{1},
Jian Gao\altaffilmark{1},
Jiaming Liu\altaffilmark{1},
Shu Wang\altaffilmark{1,2}, and
Aigen Li\altaffilmark{2}}

\altaffiltext{1}{Department of Astronomy,
                 Beijing Normal University,
                 Beijing 100875, China;
                 {\sf mengyao$\_$xue@mail.bnu.edu.cn,
                      bjiang@bnu.edu.cn,
                      jiangao@bnu.edu.cn,
                      jiaming@mail.bnu.edu.cn}
                 }

\altaffiltext{2}{Department of Physics and Astronomy,
                 University of Missouri,
                 Columbia, MO 65211, USA;
                 {\sf wanshu@missouri.edu,
                      lia@missouri.edu}
                 }


\begin{abstract}
A precise measure of the mid-infrared
interstellar extinction law is crucial
to the investigation of the properties
of interstellar dust, especially of
the grains in the large size end.
Based on the stellar parameters derived
from the SDSS-III/APOGEE spectroscopic survey,
we select a large sample of 
G- and K-type giants as the tracers of
the Galactic mid-infrared extinction.
We calculate the intrinsic stellar color excesses
from the stellar effective temperatures
and use them to determine
the mid-infrared extinction
for a given line of sight.
For the entire sky of the Milky Way
surveyed by APOGEE,
we derive the extinction
(relative to $\AKs$, the K$_{\rm S}$ band
extinction at wavelength $\lambda=2.16\mum$)
for the four \emph{WISE} bands
at 3.4, 4.6, 12 and 22$\mum$,
the four \emph{Spitzer}/IRAC bands
at 3.6, 4.5, 5.8 and 8$\mum$,
the \emph{Spitzer}/MIPS24 band at 23.7$\mum$
and for the first time, the \emph{AKARI}/S9W band
at 8.23$\mum$.
Our results agree with previous works
in that the extinction curve is flat
in the $\simali$3--8$\mum$ wavelength range
and is generally consistent with
the $\RV=5.5$ model curve except our determination
exceeds the model prediction
in the \emph{WISE}/W4 band.
Although some previous works found that the mid-IR
extinction law appears to vary
with the extinction depth $\AKs$,
no noticeable variation has been found in this work.
The uncertainties are analyzed in terms of
the bootstrap resampling method
and Monte-Carlo simulation
and are found to be
rather small.
%
%
\end{abstract}

\keywords{ISM: dust, extinction -- infrared: ISM
          -- The Galaxy: disk, ISM}


\section{Introduction}

The infrared (IR) extinction law has recently been
extensively investigated with the speedy development
of space infrared astronomy.
Based on the IR spectroscopy of the
Galactic center obtained with the
\emph{Short Wavelength Spectrometer} (SWS)
on board the \emph{Infrared Space Observatory}
(ISO), \cite{Lutz96} by the first time derived
$A_\lambda/\AKs$, the extinction at wavelengths
$\lambda$ relative to that at the $\lambda=2.16\mum$
$\Ks$ band, at several wavelengths in the
$\simali$2--20$\mum$ wavelength range from
the intensity ratios of the recombination
lines of atomic hydrogen.
Based on the ISOGAL photometric survey
around 7$\mum$ and 15$\mum$,
\cite{Jiang03} determined $A_\lambda/\AKs$
at these two wavelengths toward more than 100 sightlines
in the inner Galactic plane.
With the advent of the \emph{Spitzer Space Telescope},
much progress has been made in determining
the mid-IR extinction,\footnote{%
  In this work we refer the mid-IR to
  the $\simali$3--25$\mum$ wavelength range
  and the near-IR to
  $\simali$0.9--3$\mum$.
  }
particularly thanks to the GLIMPSE Legacy program
\citep{Churchwell2009} which obtained a rich set of
photometric data at 3.6, 4.5, 5.8 and 8.0$\mum$
with the \emph{Infrared Array Camera} (IRAC)
on board \emph{Spitzer}.
Using the photometric data from
the \emph{Spitzer}/GLIMPSE Legacy program,
\cite{Indebetouw05} calculated the extinction
in the four \emph{Spitzer}/IRAC bands for
the $l=42\degr$ and $284\degr$ sightlines
in the Galactic plane.
\cite{Flaherty07} obtained the mid-IR extinction
in the \emph{Spitzer}/IRAC bands
for five nearby star-forming regions.
\cite{Gao09} (GJL09) used the \emph{Spitzer}/GLIMPSE data
to derive the extinction in the four IRAC bands
for 131 GLIMPSE fields
along the Galactic plane within $|l|\leq65^{\rm o}$
(also see \citet{Zasowski09}).
%
\cite{Wang2013} determined the relative extinction
$A_\lambda/\AKs$ in the \emph{Spitzer}/IRAC bands
for the Coalsack nebula.
All these studies show that the mid-IR extinction
in the $\simali$3--8$\mum$ wavelength range
is much flatter than predicted by the standard
interstellar grain model for $\RV=3.1$
(\cite{WD01}; hereafter WD01)
which closely reproduces the ultraviolet (UV)
and visual extinction of the Galactic diffuse
interstellar medium (ISM).



Except the works of Lutz et al.\ (1996) and Lutz (1999)
which made use of the \emph{ISO}/SWS spectroscopy,
and the works of \cite{Boogert2011}, \cite{Boogert2013},
\cite{Chiar2007}, \cite{Chiar2011} and \cite{McClure2009},
which made use of the \emph{Spitzer}/IRS spectroscopy,
so far most of the efforts made to determine the mid-IR
extinction were based on the \emph{Spitzer}/IRAC
broadband photometry. This requires a priori assumption
that a sample of stars with the same intrinsic color indices
or with a very small dispersion can be picked up
by their locus in the color-color diagram
and/or the color-magnitude diagram.\footnote{%
%
%
  For example, \cite{Indebetouw05} selected
  a sample of red clump stars.
  GJL09 and Wang et al.\ (2013)
  considered both red clump stars and red giants.
  \cite{Zasowski09} also chose red clump stars.
  \cite{Flaherty07} assumed that the red clump stars
  and the red giant branch (RGB) stars
  in their studied field have the same mean intrinsic
  colors as that in a reference field
  which was considered as extinction-free.
  }
One then derives a statistical color-excess ratio
from the ratios of the observed colors.
Such a method has two potential sources of error:
(1) the chosen sample could be polluted
by other types of sources with mimic colors
(e.g., young stellar objects (YSOs)
and asymptotic giant stars (AGBs) appear like
red clump stars and red giants in colors); and
(2) the intrinsic color index dispersion may not
be negligibly small even among the same type of stars.
\cite{WJ14} (WJ14) found that red clump stars,
which are widely regarded to have nearly constant
IR color indices, show a dispersion of about 0.1\,mag
in the intrinsic color index $\CJKs^0$
($\equiv (J-K_{\rm S})_{\rm intrinsic}$).
%

WJ14 proposed a more accurate method to determine
the near-IR extinction law
based on the spectroscopic data from
the APOGEE survey \citep{Eisenstein2011}.
The stellar parameters
--- effective temperature and surface gravity ---
can be measured from the stellar spectroscopy,
which leads to a precise determination of
the stellar intrinsic color indices.
In the near-IR, $\CJKs^0$ can be determined
within an uncertainty less than about 0.05\,mag
if $T_{\rm eff} > 4000\,{\rm K}$
with a temperature error of 100$\,{\rm K}$.
In this work, we apply this method
to the mid-IR bands,
including the \emph{Spitzer}/IRAC bands,
the \emph{Spitzer}/MIPS 23.7$\mum$
(MIPS24 hereafter) band,
the W1, W2, W3, and W4 bands of
the \emph{Wide-field Infrared Survey Explorer}
(\emph{WISE}; Wright et al.\ 2010),
and the \emph{AKARI}/S9W band
\citep{Murakami2007,Onaka2007},
with the \emph{WISE}/W4 band and
the \emph{AKARI}/S9W band
being investigated
for the first time.

\section{Data Description}\label{DATA}

The interstellar extinction is often expressed
as reddening which is the difference between
the observed color index of an object and
its intrinsic color index.
As the absolute extinction involves the knowledge
of the absolute brightness of an object which is often
hard to determine accurately, reddening is technically
easier and more precise to measure.
In this work, we use the color excesses
to calculate the extinction.
The determination of the interstellar reddening
requires the knowledge of the intrinsic and observed colors.
In order to determine the intrinsic colors,
the stellar spectroscopic data are necessary.
Meanwhile, the broadband photometric data are necessary
for deriving the observed colors.
For these purposes, we make use of
the APOGEE spectroscopic dataset
and the \emph{AKARI/WISE/Spitzer} photometric data.


\subsection{Spectroscopic Data: The SDSS/APOGEE Survey}

The Sloan Digital Sky Survey \citep{Gunn2006}
used a dedicated 2.5\,m Ritchey-Chretien wide angle
optical telescope at Apache Point Observatory.
After the first stage of the survey, SDSS$-$I,
and the second stage SDSS$-$II, the latest third
stage started in 2008 and finished in July 2014.
APOGEE, short for
\emph{Apache Point Observatory Galaxy Evolution Experiment},
is a spectroscopic survey in the SDSS$-$III project.
APOGEE obtained  high-resolution
and high signal-to-noise ratio spectra
in the $\simali$1.51--1.70$\mum$ wavelength range
for more than 100,000 red giant stars
whose 2MASS/H band magnitudes are
between 7\,mag to 13.8\,mag
and  intrinsic color index $\CJKs^0$\footnote{%
  The $\CJKs^0$ indices given by APOGEE
  were calculated using
  the RJCE (Rayleigh-Jeans Color Excess) method
  \citep{2011RJCE} or using the extinction map of
  \cite{Schlegel98}.
  }
greater than 0.5\,mag.
The stellar parameters of the observed red giants
were determined by template matching.
The most recently released APOGEE catalog,
i.e. SDSS/DR12,  provides the stellar parameters
for about 146,000 stars \citep{2015SDSSdr12},
including the effective temperature $\Teff$,
surface gravity log\,$g$, and metallicity [M$/$H]
which are useful in determining
the stellar intrinsic colors.

\subsection{Infrared Photometric Data:
            the 2MASS, \emph{AKARI},
            \emph{Spitzer}/GLIMPSE,
            \emph{Spitzer}/MIPSGAL
            and \emph{WISE} Surveys}

\subsubsection{2MASS}

The \emph{Two Micron All Sky Survey} (2MASS) is the basis for determining the IR extinction which is often expressed as relative to the 2MASS/$\Ks$ band. In addition, the observed color index  $\CJKs$ ($\equiv  (J-K_{\rm S})_{\rm observed}$) plays a key role in choosing the zero-reddening sources. 2MASS is an all-sky survey in the near infrared bands $JH\Ks$, with a  completeness of  $>$ 99\% for $J<$15.8, $H<$15.1 and $\Ks<$14.3\,mag across the sky, except in regions of high source density. There are over 470 million sources in the 2MASS point source catalog \citep{Cutri2003}.

\subsubsection{AKARI}

\emph{AKARI} (also known as ASTRO-F)
is the first Japanese astronomical satellite
dedicated to infrared astronomical survey.
The \emph{AKARI}/S9W band (S9W hereafter) filter covers the wavelength range of 6.7--11.6$\mum$, with an effective wavelength of $\sim8.23\mum$. The 5$\sigma$ limit of the survey, 50 mJy, corresponds to a magnitude limit of 7.6\,mag with a zero magnitude flux of 56.26Jy \citep{Ishihara2010}. The cross-identification with the APOGEE/DR12 yields 1024 stars with a matching radius of 3 arcseconds. The \emph{AKARI}/S9W band is highly overlapping with the interstellar silicate feature around 9.7$\mum$, but no calculation of the interstellar extinction in this band has been well done. This work turns out to be the first of such attempt. Meanwhile, the \emph{AKARI}/L18W band is designed to cover the silicate feature around 18$\mum$. However, due to its relatively low sensitivity at $18\mum$, only 901 \emph{AKARI}/L18W sources have counterparts in the APOGEE survey. If we limit the photometry accuracy better than 0.3 mag, then only 108 sources satisfy the quality control criteria given in \S\ref{sec_quality_control}. Their observed color excess in $J-\Ks$ concentrates in the [0, 0.2] range indicating low interstellar extinction. Their small size of sample and small extension in the interstellar extinction lead us to give up the \emph{AKARI}/L18W band in following analysis.


\subsubsection{The \emph{Spitzer}/GLIMPSE Survey}

GLIMPSE (\emph{Galactic Legacy Infrared Midplane Survey Extraordinaire})
\citep{Hora2008, Churchwell2009} is a \emph{Spitzer} Legacy program.
The survey carried out imaging observation of the Galaxy in the low latitude in the four IRAC bands, centering at $3.55\mum$, $4.49\mum$, $5.73\mum$ and $7.87\mum$ respectively which are widely designated as [3.6], [4.5], [5.8] and [8.0]. The most recent GLIMPSE point source catalog from the GLIMPSE-I, II, III and 360 covers the entire GLIMPSE survey areas. The sensitivity in the four bands are 15.0, 14.5, 12.5 and 12.5\,mag  respectively at a $5\sigma$ level. Since the APOGEE catalog included the GLIMPSE photometry, in this work we simply adopt the GLIMPSE photometric data from the APOGEE catalog.



\subsubsection{The \emph{Spitzer}/MIPSGAL Survey}

The \emph{Multiband Imaging Photometer for Spitzer} (MIPS)
\emph{Galactic Plane Survey} (MIPSGAL)
is a \emph{Spitzer} Legacy program that imaged the 24$\mum$ and 70$\mum$ emission along the inner disk of the Milky Way \citep{Carey2009}, similar to the GLIMPSE project but at much longer wavelengths. Recently, a 24$\mum$ point source catalog is released to the public by \cite{Gutermuth2015}. The high quality catalog product contains 933,818 sources, with a total of 1,353,228 in the full archive catalog, which are already matched with the 2MASS, GLIMPSE, and \emph{WISE} point sources. We take advantage of this to investigate the extinction in the MIPS 24$\mum$ band ([24] hereafter).

\subsubsection{\emph{WISE}}

The \emph{Wide-field Infrared Survey Explorer} (\emph{WISE}, \citet{Wright2010}) launched in December 2009, is an infrared space telescope with a diameter of 40 cm. It performed an all-sky imaging survey  in four bands centering at 3.35, 4.60, 11.56 and 22.09 $\mum$ respectively (W1, W2, W3 and W4 bands hereafter). 
The ALLWISE catalog was released in November, 2013 and includes 747,634,026 sources. 
The AllWISE Catalog is over $95\%$ complete for sources with W1$<$17.1 mag, W2$<$15.7 mag, W3$<$11.5\,mag and W4$<$7.7 mag, achieving 16.9, 16.0, 11.5 and 8.0\,mag in W1, W2, W3 and W4, respectively \citep{2013wise}.


Table 1 lists the relevant parameters of the four photometric surveys, including the effective wavelength of each band, the area covered, the limiting magnitudes at a $5\sigma$ level, the radius and number of sources of cross-identification with the APOGEE catalog, the photometric quality and the number of sources for determining the extinction in this work which will be described in next section. As the sample sources are selected from the aforementioned five photometric surveys, the extinction derived in this work is an average over the whole surveyed sky, not for a specific line of sight.

\section{Method}\label{method}

\subsection{Determination of the Intrinsic Colors}

We determine the intrinsic color  between band $\lambda_1$ and $\lambda_2$, $C^{0}_{\lambda1\lambda2}$, of the extinction tracers from their effective temperatures measured by the APOGEE survey. The principle idea conforms to WJ14. The bluest star at a given $\Teff$ is considered to suffer no reddening so that the observed color index $C_{\lambda1\lambda2}$ will indeed be the intrinsic one. The relation of $C^{0}_{\lambda1\lambda2}$ with $\Teff$ is then determined from a series of discrete points [$\Teff$, C$^{0}_{\lambda1\lambda2}$]. On the other hand, WJ14 only dealt with the near-infrared $JH\Ks$ bands, their situation was relatively simple. In this work, multiple mid-infrared bands are involved and the data status is much more complex. Consequently, the details of the method depend on the specific situation of the relevant waveband.

\subsubsection{Control of the data quality}\label{sec_quality_control}

The APOGEE catalog includes A-, F-, G- and K-type stars. We choose only the G-, K-type giants. The G- and K-type giants are the dominant members of the APOGEE catalog and have reliably determined stellar parameters. They are bright enough in the infrared to probe deep infrared extinction. Furthermore, exclusion of other types reduces the range of $\Teff$ to improve the analytical fitting of its relation with $C^{0}_{\lambda1\lambda2}$.

The adopted data quality control is as following:
\begin{enumerate}
  \item The spectral type is `GK' and log\,$g<$3. This selects G- and K-type giant stars. However, it should be noted that the stellar spectral libraries used to determine the stellar parameters in the ASPCAP pipeline of APOGEE combine `G-type' and `K-type' into one class, `GK' class\footnote{http://www.sdss.org/dr12/irspec/aspcap/}.
  \item Metallicity [M$/$H]$>$-1. The metallicity is the secondary parameter which affects the intrinsic colors, but its effect is much weaker in the infrared than in the visible \citep{Ramirez05}. This criterion does not really exclude metal-poor stars, but excludes those with significant uncertainty of measured metallicity because the determination of stellar parameters for metal-poor stars is more challenging as their spectral lines are weaker. There is no attempt to distinguish metal-rich and metal-poor stars in this work, and we will carry out a separate, independent analysis of the metallicity. WJ14 pointed out that the metallicity effect is not clear because the metal-poor sample in the APOGEE survey is not big enough.
  \item VSCATTER $<$0.3$\km\s^{-1}$.
        With a claimed velocity accuracy of
        $\simali$0.1$\km\s^{-1}$ of APOGEE,
        this criterion corresponds to a 3-sigma level
        to exclude binary stars whose colors
        must be different from single stars.
  \item The spectrum's SNR$>$100 and the photometric error
        of the $J$-band and $\Ks$-band less than 0.05\,mag.
\end{enumerate}

Under these criteria, 63,330 sources are picked up to form the sample for subsequent cross-identification with photometric surveys.

For the different photometric surveys we used, the photometric accuracies vary. In order to guarantee the significance of statistics and the quality of photometry, the constraint on the photometric error differs from survey to survey. Namely, the photometric accuracy is required to be better than 0.1mag for the \emph{WISE} and GLIMPSE surveys, and 0.2\,mag for the \emph{AKARI} and MIPSGAL surveys. Moreover, the radius of cross-identification with APOGEE also differs. The details of cross-identification and photometric quality are listed in Table \ref{data quality}, as well as the number of stars cross identified with the selected APOGEE sample. The sample with photometric quality control and identified in the selected APOGEE sample forms the basis for our analysis of the intrinsic color indices and the interstellar extinction.

\subsubsection{Definition of the blue edge in the $\Teff$ vs $\CJKs$ diagram}

The bluest stars in the $\Teff$ vs color index diagram are considered as non-reddened ones as originally suggested by \cite{Ducati2001}. 
The feasibility of this method relies on the condition that the sample does include non-reddened objects. As long as the non-reddened stars are included in the sample, they should show the bluest observed colors among the same type of stars and their observed colors are considered to be the intrinsic colors of this type of stars.

\begin{figure}
\centering
\includegraphics[scale=0.6]{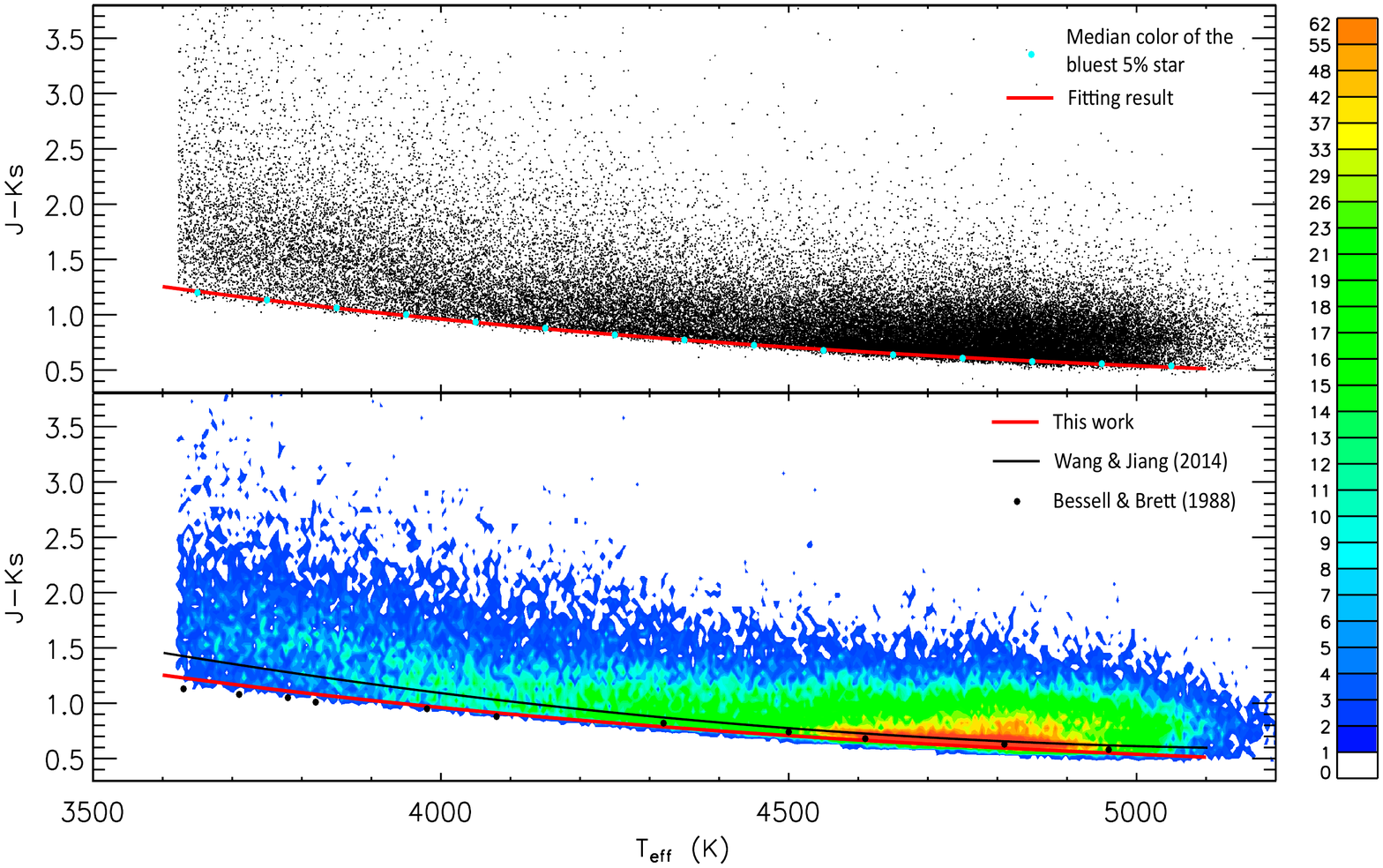}
\caption{
          Distribution of the selected APOGEE stars in the observed color index $\CJKs$ vs. the effective temperature $\Teff$ diagram. The upper panel displays the colors of all the sample stars (black dots), the median colors of the 5\% bluest stars (cyan dots), and the line of a quadratic fitting (red line). The bottom panel shows the density contour of the upper panel, together with the fitting line of WJ14 (black solid line) and the results of Bessel \& Brett (1988) (black solid circles).
\label{fig2}}
\end{figure}

The $\Teff$-$\CJKs$ diagram of the selected 63330 APOGEE stars is shown in Figure~\ref{fig2}. The blue edge is very apparent at a first glance. But the mathematical definition of the blue edge is not straightforward. We define the blue edge in two steps. First, the median color of the bluest 5$\%$ stars in a bin of $\Delta \Teff$=100K is chosen as the typical value of $\CJKs^0$ in this $\Teff$ bin.
Then a function, either exponential or quadratic, is used to fit the typical values. Since this method is completely mathematical, the form of the function has no physical meaning. Which function is selected depends only on the goodness of fitting. As for $\CJKs^0$, an exponential function fits better.




$C^{0}_{\rm JKs}$ derived in this way is compared with the work of \cite{Bessell88} and WJ14. A quantitative comparison is presented in Table \ref{Teffcompare}  at some typical temperatures.  In Figure~\ref{fig2}, it can be seen that this color index agrees quite well with the \cite{Bessell88} values, with the difference mostly less than 0.05 mag, slightly redder at low $\Teff$ and slightly bluer at high $\Teff$. On the other hand, a systematically bluer color than that of WJ14 is found. This is probably because that WJ14 may have over-estimated this parameter, particularly at the low temperature end. As they confined the sources to be within the Galactic plane with $|b|<5\degr $ where the extinction is very severe, there could be no giants with zero-reddening, even the bluest ones in their samples cannot avoid being reddened. Meanwhile, this work does not put any constraint on the Galactic latitude, and some high latitude stars with no or little extinction are present and displayed as the bluest in the $\Teff$-color index diagram. Because low temperature red giants are usually brighter and further, they suffer more extinction and would appear redder than their intrinsic colors. This is consistent with the increasing discrepancy to the low temperature end between the present work and WJ14.

\subsubsection{Selection of stars of zero-reddening}

The reddening free stars are selected based on the $\Teff$-$\CJKs$ diagram. Once the blue edge is defined, the stars close to the fitting line are regarded as zero-reddening stars. The ``close-to'' distance is defined as that the star deviates from the fitting line (i.e. the red line in Figure~\ref{fig2}) not larger than the uncertainty ($\equiv \sqrt{J^2_{\rm err}+\Ks^2_{\rm err}}$) of the observed color index $\CJKs$. That is to say, the deviation from the line of intrinsic color can be completely attributed to the photometric error other than the interstellar extinction.

\begin{figure}
\centering
\includegraphics[scale=0.6]{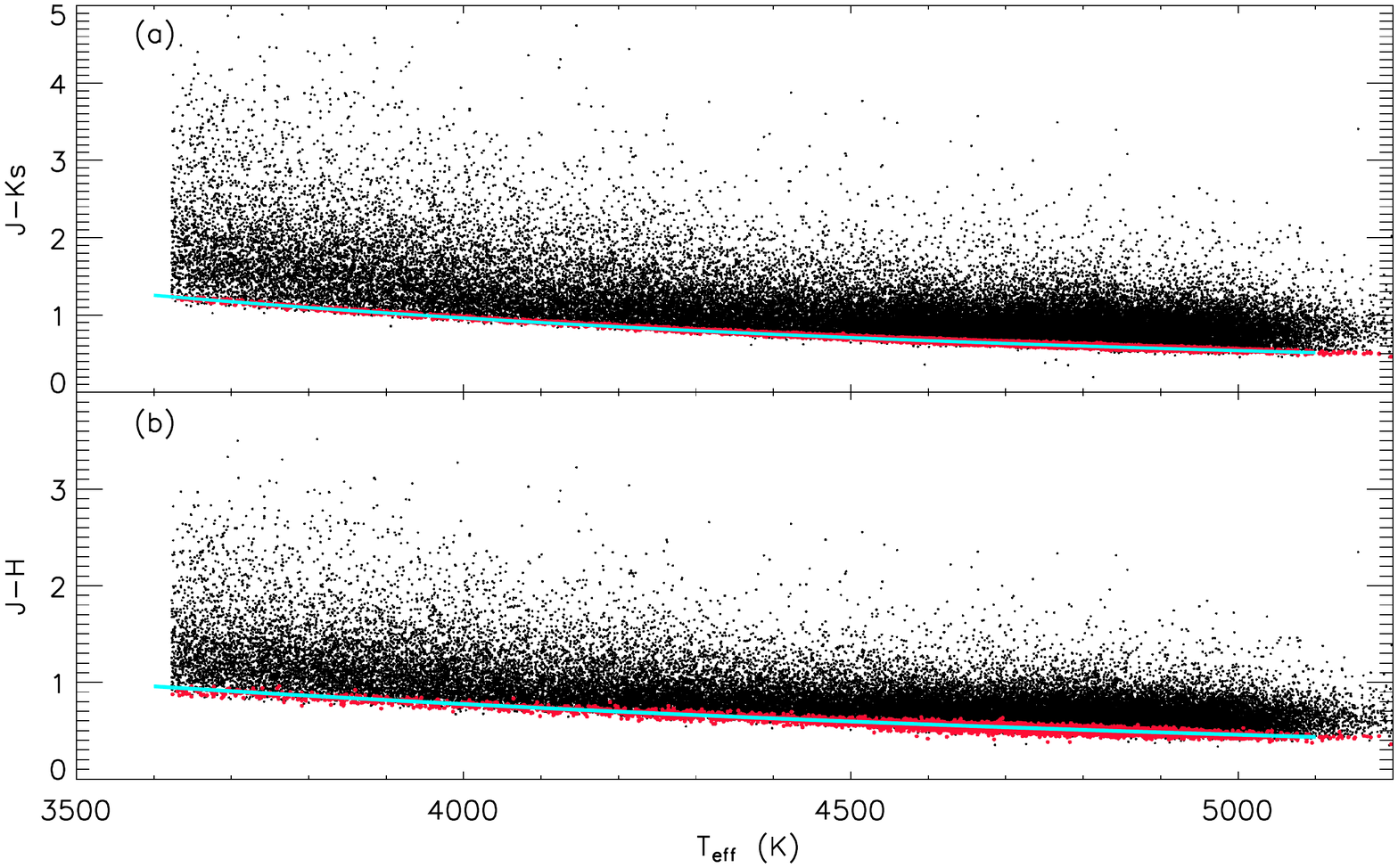}
\caption{
         Determination of the relation of the color index $C_{\rm {JH}}^0$ with $\Teff$. In the upper panel, the result of fitting $C_{\rm {J\Ks}}^0$ with $\Teff$ from Figure~\ref{fig2} is shown by a cyan solid line, and the zero-reddening stars are decoded by red dots. In the bottom panel for the color index $C_{\rm {JH}}^0$, the fitting result of the zero-reddening stars (red dots) is delineated by a cyan solid line.
\label{fig4.1}}
\end{figure}

The method of selecting zero-reddening stars works very well in most of the wavebands. For example, the method fits the color $C_{\rm JH}$ very well as shown in Figure~\ref{fig4.1}. In addition, for the first three \emph{WISE} bands, the bluest stars among the sample cross-identified between the APOGEE and ALLWISE catalogs shown in Figure~\ref{fig41}(a), Figure~\ref{fig42}(a), and Figure~\ref{fig43}(a), are closely consistent with the fitting line derived from the APOGEE data. Moreover, they are unanimously the bluest stars in the $\Teff$ vs. $\Ks-W1/W2/W3$ diagrams. This consistency is expected from the fact that the true zero-reddening stars should have zero-reddening in any color index because reddening is of interstellar origin rather than stellar origin. 

In spite of the consistency in most of the $\Teff$-color index diagrams, there appear to be some deviations as well. A typical case occurs in the $\Teff$ vs. $\Ks-S9W$ diagram. The zero-reddening stars selected from the closeness of their observed $\CJKs$ to the fitting line do not look like the blue edge in the $\Teff$ vs. $\Ks-S9W$ diagram. We believe this inconsistent appearance is not true. In fact, the photometric error in the $AKARI$/S9W band is relaxed to 0.2\,mag for the relatively low sensitivity in this band in order to guarantee the size of the sample. The average deviation from  fitting the zero-reddening stars (the blue line in Figure~\ref{fig3}) is 0.16, which is comparable to the photometric uncertainty. The same reason can account for the dispersion in the $\Teff$ vs. $\Ks$-W4 diagram.

The four IRAC bands and the [24] band exhibit another type of disagreement. As can be seen in Figure~\ref{fig31}(a), Figure~\ref{fig32}(a), Figure~\ref{fig33}(a), Figure~\ref{fig34}(a), Figure~\ref{fig35}(a), all the stars with the IRAC or MIPS measurement lie above the line (the blue solid line) delineating the relation between $\Teff$ and C$^{0}_{\rm JKs}$. This means that all the sample stars are redder than the intrinsic colors. It can be recalled that the discrepancy between this work and WJ14 is explained by the fact that WJ14 only took the low-latitude stars  and could not avoid interstellar extinction. The same reason can explain the disagreement of the bluest stars with the expected intrinsic color indices in the four \emph{Spitzer}/IRAC bands and \emph{Spitzer}/[24] band. The GLIMPSE and MIPSGAL surveys targeted only the Galactic plane with the latitude smaller than 5\degr~and all the targets suffer some extent of extinction. To proceed in this case, a red-ward shift of the intrinsic color index is added in order to choose a sample of faked zero-reddening stars. For the four IRAC bands, a 0.17\,mag shift in $\CJKs$ was adopted, which is roughly equivalent to $\AV\sim1$\,mag (assuming $A_{\rm J}/\AV=0.282$, $\AKs/\AV=0.112$ according to \cite{Rieke85}); for the [24] band, stars experience more interstellar extinction, a shift of 0.21\,mag is added to obtain $C_{\rm \Ks[24]}$. The faked zero-reddening stars are used to fit the relation between $\Teff$ and $C_{\rm \Ks[3.6]}$,  $C_{\rm \Ks[4.5]}$,  $C_{\rm \Ks[5.8]}$, $C_{\rm \Ks[8.0]}$, and $C_{\rm \Ks[24]}$, which are the blue dash line in the lower panel of the diagrams. Of course, the color fitted in this way is redder than the true intrinsic color index. But this shift only affects the intercept and does not affect the slope when determining the color-excess ratio $E_{\rm \Ks\lambda}/E_{\rm{J\Ks}}$ (color excess $E_{\lambda1\lambda2} \equiv C_{\lambda1\lambda2}-C_{\lambda1\lambda2}^0$ ) in the linear fitting, which will be discussed in Section 3.3. To compensate for this red-ward shift, a blue-ward shift is performed. The blue-ward shift values are determined by the color-excess ratio $E_{\Ks\lambda}/E_{\rm{J\Ks}}$ from our calculation,  which are 0.26, 0.31, 0.36, 0.33, 0.43 for [3.6], [4.5], [5.8], [8.0] and [24] respectively.

\subsubsection{The $\Teff$-$C_{\lambda1\lambda2}^0$ relation}\label{sec_fit_ic}

With the zero-reddening sample defined in the way described above, an analytic expression is derived for the $\Teff$-$C_{\lambda1\lambda2}^0$ relation. This is done  in each photometric band by fitting a function, either exponential, quadratic or linear, to the zero reddening stars. For most of the bands, a quadratic function fits the relation best. However, for $C^{0}_{\rm JH}$, an exponential function fits better, while for $C^{0}_{\rm Ks[24]}$, a linear function works better mainly due to the smallness of the zero-reddening sample.

The intrinsic color indices as a function of $\Teff$ derived in this way are given as following, and
the values for $\Teff$ from 3600\,K to 5200\,K with a step of 100\,K are listed in Table \ref{tab_colors}.

\begin{equation}\label{JK_Teff}
C^0_{\rm{J\Ks}}=20.285\times{\rm exp}\left(-\frac{\rm{T_{eff}}}{1214{\rm K}}\right)+0.209
\end{equation}

\begin{equation}\label{JH_Teff}
C^0_{\rm{JH}}=6.622\times{\rm exp}\left(-\frac{\rm{T_{eff}}}{1846{\rm K}}\right)+0.019
\end{equation}

\begin{equation}\label{K9_Teff}
C^0_{\rm{\Ks S9W}}=0.087\times{\Teffqk}^2-0.869\times{\Teffqk}+2.326
\end{equation}

\begin{equation}\label{KW1_Teff}
C^0_{\rm{\Ks W1}}=0.026\times{\Teffqk}^2-0.268\times{\Teffqk}+0.755
\end{equation}

\begin{equation}\label{KW2_Teff}
C^0_{\rm{\Ks W2}}=0.018\times{\Teffqk}^2-0.115\times{\Teffqk}+0.134
\end{equation}

\begin{equation}\label{KW3_Teff}
C^0_{\rm{\Ks W3}}=0.079\times{\Teffqk}^2-0.736\times{\Teffqk}+1.805
\end{equation}

\begin{equation}\label{KW4_Teff}
C^0_{\rm{\Ks W4}}=0.105\times{\Teffqk}^2-1.050\times{\Teffqk}+2.770
\end{equation}

\begin{gather}\label{K36_Teff}
C^0_{\rm{\Ks [3.6]}}
=0.017\times{\Teffqk}^2-0.230\times{\Teffqk}+0.753
\end{gather}

\begin{gather}\label{K45_Teff}
C^0_{\rm{\Ks [4.5]}}
=-0.075\times{\Teffqk}^2+0.682\times{\Teffqk}-1.605
\end{gather}

\begin{gather}\label{K58_Teff}
C^0_{\rm{\Ks [5.8]}}
=0.017\times{\Teffqk}^2-0.221\times{\Teffqk}+0.696
\end{gather}

\begin{gather}\label{K80_Teff}
C^0_{\rm{\Ks [8.0]}}
=0.063\times{\Teffqk}^2-0.583\times{\Teffqk}+1.414
\end{gather}

\begin{gather}\label{K24_Teff}
C^0_{\rm{\Ks [24]}}
=-0.078\times{\Teffqk}+0.406
\end{gather}


All the bands conform to the law that the color index rises with decreasing $\Teff$ although the amount differs from band to band. However, there are two exceptions, i.e., IRAC/[4.5] and \emph{WISE}/W2. According to Table \ref{tab_colors}, $C_{\rm {\Ks W2}}^0$ presents a completely inverse tendency in this range of $\Teff$, i.e. rising with $\Teff$. $C_{\rm \Ks[4.5]}^0$ rises with $\Teff$ from 3600\,K to 4500\,K, and then falling from 4600\,K to 5200\,K. Moreover, these two color indices are mostly negative, which implies that the star is brighter in $\Ks$ band than in the W2 or [4.5] band in terms of magnitude. According to our selection criteria, the sample stars are G- or K-type stars, significantly redder than A0-type stars -- standard of zero color index.  In fact, the effective wavelength of W2 (4.60$\mum$) is very close to that of [4.5] (4.49$\mum$, see Table \ref{tab_colors}). The difference between $C_{\rm \Ks W2}^0$ and $C_{\rm \Ks[4.5]}^0$ is on the order of 0.05 mag, which confirms the internal consistency and manifests the reliability of the tendency and the negative color index. The possible reason is that some absorption band(s) around 4.5$\mum$  weakens the W2 and [4.5] band and that the absorption band(s) strengthen with decreasing $\Teff$. \cite{Bernat1981} found that some red giants have circumstellar envelopes rich in CO gas that absorbs at 4.6$\mum$. In Galactic and extragalactic dense clouds, various ice features have been observed around this band, including a number of weaker features at 4.27$\mum$ due to the C-O stretching mode of CO$_2$ ice, a feature at 4.67$\mum$ due to the C-O stretching mode of CO ice. As for these G- and K-type giant stars, ice should be lacking because their $\Teff$ is still high and mass loss rate is still very low in comparison with AGB stars. The absorption is probably caused by CO gas. Interestingly, the [4.5] extinction was previously found to be significantly higher than the other three IRAC bands, which was attributed to the CO absorption in this band in their extinction tracers, i.e. giant stars \citep{gao2013}. This work confirms their suspicion. On the other hand, how the absorption changes with $\Teff$ and affects the W2 and [4.5] bands differently needs further investigation.
It would be interesting to
note that the \emph{Spitzer}/IRAC [8.0] band
and potentially the \emph{AKARI}/S9W band as well
could also be affected by the photospheric SiO
gas absorption around 8$\mum$
\citep{Boogert2013,Boogert2015}.
But such an effect is not apparent
in the derived intrinsic colors related to
the \emph{Spitzer}/IRAC [8.0] or \emph{AKARI}/S9W bands.
Probably the SiO molecular band is not strong
or common enough to
appreciably affect these broad band photometries.

\subsection{Determination of the Near-IR Extinction Law}\label{sec_nir_ext}
The near-IR extinction law was already derived by WJ14 in the same way as in the present work. But we still re-calculate the ratio between the color excesses $E_{\rm JH}$ and $E_{\rm J\Ks}$. The main reason is that WJ14 constrained their sample to be within the Galactic plane with unavoidable extinction which led to an over-estimation of the intrinsic color indices. Additionally, SDSS/DR12 improved the calculation of stellar parameters and enlarged the catalog. With the newly determined intrinsic color indices based on the updated APOGEE catalog, it is found that $E_{\rm JH} \approx 0.652\times E_{\rm J\Ks} - 0.008$ with a ultra tight linear correlation and very little uncertainty. The color-excess ratio, 0.652, is closely consistent with the WJ14 result 0.64.

The near-IR extinction is commonly expressed by a power law $A_\lambda \propto \lambda^{-\alpha}$. Given $E_{\rm JH} / E_{\rm J\Ks} = 0.652$, the power law index $\alpha$ was derived to be 1.79 when adopting $\lambda_J=1.235\mum$, $\lambda_H=1.662\mum$, $\lambda_\Ks=2.159\mum$ as the effective wavelength of the 2MASS $JH\Ks$ photometry system \citep{Cohen2003}. 
Consequently, the extinction of the $J$ to $\Ks$ band is  $A_{\rm J}/A_{\rm \Ks}=(\frac{\lambda_{\rm \Ks}}{\lambda_{\rm J}})^\alpha\approx2.72$, which will be used in following derivation of the extinction in other bands. Although the power law provides a convenient way to calculate the extinction, the index $\alpha$ and the resultant $A_{\rm J}/A_{\rm \Ks}$ are very sensitive to the adopted wavelength of the $JH\Ks$ bands. In contrast, the color-excess ratio is a more stable and reliable description of the extinction law.



\subsection{Determination of the Mid-IR Extinction Law}\label{sec_method_f}
The stellar color excess can be calculated in a straightforward way once the intrinsic color index is derived from its effective temperature for each star in the selected sample. In principle, the derived color-excess ratio for one star can represent the extinction law. However, the uncertainty in stellar parameters and photometry will lead one to question about the reliability of an individual star. The large sample resulting from the cross-identifications of the photometric surveys and APOGEE spectroscopic survey makes it feasible to perform a reliable statistical analysis. Therefore, the color-excess ratio between $E_{{\rm J\Ks}}$ and $E_{\rm \Ks\lambda}$ is determined by a statistical linear fitting to the cross-identified sample.

The extinction at $\lambda_x$ relative to $\Ks$ band $A_{\lambda_{x}}/\AKs$ can be derived by $k_{x}$ (the linear fitting coefficient between $E_{\rm \Ks\lambda}$ and $E_{{\rm J\Ks}}$):

\begin{equation}\label{slope}
k_{x}\equiv\frac{E_{\rm \Ks\lambda_x}}{E_{{\rm J\Ks}}}
=\frac{C_{\rm \Ks\lambda_x}-C^0_{\rm \Ks\lambda_x}}{\CJKs-\CJKs^0}
=\frac{A_{\rm \Ks}-A_{\lambda_x}}{A_{\rm J}-A_{\rm \Ks}}
\end{equation}

\begin{equation}\label{eq_ext}
A_{\lambda_{x}}/A_{\rm \Ks} = 1 + k_{x}\left(1 - A_{\rm J}/A_{\rm \Ks}\right) ~~.
\end{equation}

Eq.\,(\ref{eq_ext}) tells that $A_{\lambda_{x}}/\AKs$ depends on the adopted value of $A_{\rm J}/A_{\rm \Ks}$. A larger $A_{\rm J}/\AKs$ would yield a smaller $A_{\lambda_{x}}/\AKs$ since $k_{x}$ is positive. We adopted $A_{\rm J}/\AKs=2.72$ as derived in Section \ref{sec_nir_ext}.


\subsection{Error Analysis}

The statistical uncertainty of our linear fitting is analyzed in two ways. One is the bootstrap resampling method and the other is the Monte-Carlo simulation. The bootstrap resampling method, introduced by \cite{Efron79}, generates a large number of datasets, each with the same number of data points randomly drawn from the original sample. Each drawing is made from the entire dataset and some points may be duplicated while some may be missing. We carried out 20,000 times bootstrap resampling for each pair of color excesses to analyze the uncertainty. On the other hand, the Monte-Carlo simulation constructs new sample by taking  the photometric error into account. We also carried out 20,000 times simulation of the samples.

The distribution of the slope of the bootstrap resampling is shown for the color-excess ratio $E_{\rm \Ks S9W}/E_{\rm {J\Ks}}$ in Figure~\ref{BootstrapE9} as an example. The distribution is well fitted by a Gaussian function with a peak at 0.274 and width of 0.020. In comparison, our linear fitting yields the ratio of 0.273, which is very consistent with the bootstrap resampling. The intercept is also highly consistent, both the bootstrap resampling and our linear fitting got -0.013. The Monte-Carlo method re-confirms our fitting. On the uncertainty of linear fitting, the bootstrap resampling yields a dispersion of 0.02 in the slope and 0.006 in the intercept. The Monte-Carlo simulation results in the same order of magnitude uncertainty. Both statistical methods show that our linear fitting has a very high precision, even in the \emph{AKARI}/S9W band which has the largest photometric error and smallest sample among all the bands studied. The results of statistical analysis are presented in Table \ref{ST_EE}.

The result on the \emph{WISE}/W4 band needs to be cautiously treated. The ratio $E_{\rm \Ks W4}/E_{\rm J\Ks}$ derived from the bootstrap resampling is 0.371 while it is 0.360 from the Monte-Carlo simulation. The discrepancy between these two simulations reaches 0.01. Such discrepancy is one order of magnitude bigger than that of other bands. Similar to the \emph{AKARI}/S9W band, the increase of the uncertainty can be attributed to the relatively low sensitivity in the \emph{WISE}/W4 band (its 5 $\sigma$ detection limit is 8.0 mag) and the small number (2008) of sample after cross-identifying with the APOGEE survey.

These statistical analyses only consider the photometric error. Meanwhile, there also exists some uncertainty in the determination of the intrinsic color index. The uncertainty of the intrinsic color index arises from the error of the effective temperature in addition to the photometric error. As an example, an error of 100\,K in $\Teff$ would result in an error of less than 0.05\,mag for the near-infrared color index $\CJKs^0$. This is comparable to the photometric error. Taking the effective temperature error, the uncertainty in the linear fitting will be about 1.4 times the tabulated value in Table \ref{ST_EE}.


\section{Results and Discussion}

\subsection{Relative extinction in the mid-IR bands}

Based on the selected data and using the method described above, the extinction relative to $\Ks$ band is derived for the mid-infrared bands.
Figure~\ref{JK_KW1234}, \ref{JK_KIRAC} and \ref{JK_K9_24} show the linear fitting of $E_{\rm \Ks\lambda}$ with $E_{{\rm J\Ks}}$.
The values of $A_\lambda/\AKs$ are listed in Table \ref{EAresult} for $A_{\rm J}/\AKs=2.72$ (i.e., $A_\lambda \propto \lambda^{-1.79}$, see \S\ref{sec_nir_ext}). For comparison, the case for $A_{\rm J}/\AKs=2.52$ of \cite{Rieke85} which corresponds to $A_\lambda \propto \lambda^{-1.65}$ is also calculated.

The extinction in the \emph{Spitzer}/IRAC bands have been studied widely and the reported results are broadly consistent with each other in the sense that they all show a flat extinction law in the $\simali$3--8$\mum$ wavelength range, although there does exist some slight discrepancies. The results of \citet{Indebetouw05} are basically the lower limit, while that of GJL09 are the upper limit. Figure~\ref{extcurve} shows that the result of present work yields even smaller values in all four IRAC bands than that of \citet{Indebetouw05}. Indeed, this lower value comes from our adoption of $A_{\rm J}/\AKs=2.72$. Previously, \citet{Indebetouw05} and GJL09 both adopted $A_{J}/\AKs=2.52$. With $A_{\rm J}/\AKs=2.52$, our result would lie between that of \citet{Indebetouw05} and GJL09 (cf. Table \ref{EAresult}). In general, our result is consistent with previous work, in particular that the extinction between 3--8$\mum$ is clearly higher than the WD01 $\RV=3.1$ curve. 

The extinction in the \emph{WISE} bands has previously been partly studied. \cite{Yuan2013} obtained the relative extinction in the W1 and W2 bands based on a large-scale photometry. They derived $A_{\rm W1}/\AKs=0.63$ and $A_{\rm W2}/\AKs=0.50$ for non-galactic plane regions. \cite{Davenport2014} also based on photometry and derived $A_{\rm W1}/\AKs=0.60$, $A_{\rm W2}/\AKs=0.33$ and $A_{\rm W3}/\AKs=0.87$ for the galactic latitude $|b|>10\degr$ region. Their results were consistent in the W1 band, while apparently discrepant in the W2 band. Our work yields $A_{\rm W1}/\AKs=0.591$, $A_{\rm W2}/\AKs=0.463$, $A_{\rm W3}/\AKs=0.537$ and $A_{\rm W4}/\AKs=0.364$ with the latter derived for the first time. We note that a direct comparison between our result with that of \cite{Yuan2013} and \cite{Davenport2014} may not be very meaningful as we adopted $A_{\rm J}/\AKs$=2.72 while \cite{Yuan2013} assumed $A_{\rm J}/\AKs$=2.38 (see \cite{ChenBQ2014}) and \cite{Davenport2014} assumed $A_{\rm J}/\AKs$=2.0.

As for the \emph{Spitzer}/[24] band, \cite{Flaherty07} obtained $A_{\rm {[24]}}/\AKs=0.46$ for star-forming regions. Similarly, \cite{Chapman2009} studied $A_{\rm {[24]}}/\AKs$ for a few molecular clouds, and got different values towards different optical depth of extinction. For the largest extinction, i.e. with $\AKs \geq 2.0$,  $A_{\rm [24]}/\AKs = 0.34$. This value increases towards smaller extinction sightlines. In fact, to lower extinction sightlines, the weakness of the extinction at this wavelength would make it hard to measure the extinction accurately. So we only compare that of the sightline with the largest extinction. It can be seen from Figure~\ref{extcurve}, that this work yields a smaller extinction in the \emph{Spitzer}/[24] band, with a value of $A_{\rm [24]}/\AKs \approx 0.264$. 

For the first time, we derived the extinction in the \emph{AKARI}/S9W band. This band centers within the silicate feature around 9.7$\mum$ and becomes important to delineate the profile of extinction. The derived value of $A_{\rm S9W}/\AKs$ is 0.530, higher than the IRAC [8.0] band. Although this band suffers relatively large uncertainty, it agrees very well with the profile defined by ``astronomical'' silicate of the WD01 $\RV=5.5$ interstellar dust model.

\subsection{The Silicate-Related Bands}
Previously, \cite{McClure2009} and \cite{Boogert2013} described the profile of the silicate 9.7$\mum$ feature toward a series of star-forming regions and Lupus cloud using \emph{Spitzer}/IRS spectroscopy. As for the extinction calculated from photometric data, only the IRAC [8.0] band centering at 8$\mum$ was available, which generally lies above the WD01  dust model curve of $\RV=5.5$. In this work, the extinction in the \emph{AKARI}/S9W and \emph{WISE}/W3 bands is derived and adds new constraint on the 9.7$\mum$ profile. Together with the newly determined value at the IRAC [8.0] band, the result agrees very well with the WD01 dust model curve of $\RV=5.5$. The extinction derived here for the \emph{AKARI}/S9W, \emph{WISE}/W3 and IRAC [8.0] bands appears to be consistent with the 9.7$\mum$ absorption feature of the \cite{DraineLee1984} astronomical silicate but appreciably broader than that of \cite{Chiar2006}.

The broad band photometric extinction around the 20$\mum$ silicate absorption profile is poorly known. The \cite{Lutz99} result toward the Galactic center shows no decreasing of extinction at all from 12$\mum$ to 20$\mum$, which is apparently above the prediction of the WD01 dust model. As mentioned above, \cite{Chapman2009} and \cite{Flaherty07} obtained an extinction in the \emph{Spitzer}/[24] band, also much higher than the $\RV=5.5$ model curve. However, in the \emph{Spitzer}/[24] band, this work is closely consistent with the WD01 dust model curve of $\RV=5.5$. Meanwhile, the extinction of the 22$\mum$ \emph{WISE}/W4 band is much higher than the model curve. Although the sample of W4 band has more (2008) stars than the [24] sample (806), the [24] sample has more stars with large color excess $E_{\rm \Ks[24]}$ than the W4 band sample. As shown in Figure~\ref{JK_KW1234} and Figure~\ref{JK_K9_24}, a reasonable portion of the [24] sample has $E_{\rm J\Ks} > 1.0$ while only very few W4 stars are so reddened,  which can be explained by the fact that MIPSGAL surveyed the Galactic plane with deep extinction while \emph{WISE} surveyed the entire sky, much of which is high-latitude low-extinction area. The result at [24] may be more reliable due to its larger optical depth. But the error analysis shows that the uncertainties in these two bands are comparable. A better method is needed to characterize the profile in this wavelength range.

One may argue that the derived silicate extinction profile might be contaminated by the circumstellar silicate grains which may be present around oxygen-rich evolved giants. To this end, we checked the objects used as tracers of extinction. As the \emph{AKARI}/S9W and \emph{WISE}/W3 band would be most affected, the sources with circumstellar silicate features shall display excess in these two bands. A cross-identification in these two bands is made. Their color-color diagram of $\Ks-W3$ vs. $J-\Ks$, Figure~\ref{JK_KW3_liu_xue}, manifests that sources which could have silicate features always show excess in the W3 band. On the other hand, our sample stars (purple dots in the diagram), selected APOGEE GK-type giants, mostly do not show excess in the W3 band. By further excluding the 3-sigma outliers, we see that the stars used for determining extinction law (blue dots in the diagram) have no excess in the W3 band. Therefore, there is no need to worry about the influence of circumstellar silicates on the accuracy of the extinction determination around the silicate features.

\subsection{Differentiation in extinction depth}

It has been long known that the extinction law changes with interstellar environments in the ultraviolet and visible, which can be characterized by the total-to-selective extinction ratio parameter $\RV$ \citep{Cardelli89}. In the near-infrared, there are arguments both against and for the variation of the extinction law (see WJ2014 and references therein), while WJ2014 claimed that the extinction law in the $JH\Ks$ bands is universal. Based on the study of three molecular clouds of the extinction law from 3.6 to 24$\mum$, \citet{Chapman2009} found the extinction law changing with the total line-of-sight extinction. Specifically, the relative extinction $A_\lambda/\AKs$ increases with $\AKs$ in the four \emph{Spitzer}/IRAC bands, while decreases in the \emph{Spitzer}/MIPS 24$\mum$ band. Working in a very similar wavelength range (5--20$\mum$), \citet{McClure2009} found that the relative extinction $A_\lambda/\AKs$ also increases with $\AKs$, but the tendency is kept until 20$\mum$ which is the long end of the wavelength range of her study with the \emph{Spitzer}/IRS data. It seems that these two works agree with each other in the 3--8$\mum$ range, while are inconsistent for longer wavelength.
However, \cite{Roman-Zuniga2007}
and \cite{Ascenso2013} found no evidence
for the dependence of the mid-IR extinction law
on the total dust extinction.
As summarized in \cite{WLJ14},
the flat mid-IR extinction is not only seen in dense regions,
but also in diffuse clouds
(e.g., the low-density lines of sight in the Galactic midplane
[see Zasowski et al.\ 2009], and the diffuse and translucent regions
of the Coalsack nebula [see \cite{Wang2013}]).

The data in this work covers a wide range of extinction, providing the possibility to study the variation of extinction law with the depth of extinction. For the convenience of comparison, the data are divided into three groups: $E_{\rm{J\Ks}} < 0.86$, $0.86 \leq E_{\rm{J\Ks}} < 1.72$ and $E_{\rm{J\Ks}} \geq 1.72$. When converted to $\AKs$, the division becomes $\AKs < 0.50$, $0.50 \leq \AKs < 1.0$ and $\AKs \geq 1.0$ with $A_{\rm J}/\AKs=2.72$ derived in this work, which coincides with the division of \citet{Chapman2009}, but no separate division is considered for $\AKs \geq 2.0$ because the sub-sample would be too small to have reliable result.

The division into sub-sample is based on the sample selected according to the criteria described in \S\ref{sec_quality_control}, including the qualities of photometry and stellar parameters, as well as the deviations less than 3-sigma in the linear fitting of color excess $E_{\rm \Ks\lambda}$ versus $E_{\rm{J\Ks}}$. A linear fitting is performed to the sub-samples of stars in different range of $E_{\rm{J\Ks}}$, with a condition that the line is forced to pass the point (0,0). The results of linear fitting are listed in Table \ref{binEresult}. Also listed are the number of objects used in the linear fitting, and the scattering of fitting expressed by the standard deviation defined as $\sqrt{\Sigma(y_{\rm fit}-y_{\rm i})^2/(N-1)}$ where $y_{\rm fit}$ is the $E_{\rm \Ks\lambda}$ value from the linear fitting, $y_{\rm i}$ is the $E_{\rm \Ks\lambda}$ data and N is the number of sources used in the fitting. It should be noted that no value is obtained for $E_{\rm{\Ks W4}}/E_{\rm{J\Ks}}$ for $E_{\rm{J\Ks}} > 1.72$ as there is only one source. Taking $A_{\rm J}/\AKs=2.72$, the color excess ratios are converted into the relative extinction $A_{\rm \lambda}/\AKs$ in Table \ref{binAresult}.

The variation of the relative extinction $A_{\rm \lambda}/\AKs$ with $\AKs$ is very small in the bands for which the results are most reliable, i.e. the first three \emph{WISE} bands W1, W2 and W3, and the four \emph{Spitzer}/IRAC bands, [3.6], [4.5], [5.8] and [8.0]. The dispersion falls mostly below 0.02. The \emph{Spitzer}/MIPS [24] and \emph{AKARI}/S9W bands do neither show any noticeable variation although their uncertainty are bigger due to their smaller sample sizes. The only band that shows appreciable variations with $\AKs$ is the \emph{WISE}/W4 band, from 0.556 to 0.398 when $\AKs$ increases from [0,\,0.5] to [0.5,\,1]. However, with the uncertainty taken into account, this variation is within the determination error and may not be real. Therefore, we conclude that our result doesn't show considerable variation of $A_{\rm \lambda}/\AKs$ with $\AKs$.

Our conclusion of no appreciable variation of $A_{\rm \lambda}/\AKs$ with $\AKs$ differs from \citet{Chapman2009} and \citet{McClure2009}, as shown in Table\,\ref{binAresult} and Figure~\ref{bin_curve}. This apparent discrepancy may not be true if the errors are taken into account. On the other hand, both \citet{Chapman2009} and \citet{McClure2009} targeted star-forming regions which are dense enough for grain growth to occur. In contrast, this work does not distinguish interstellar environments and our result is smoothed into an average even though some sightlines towards dense clouds are included.
Nevertheless, using the \emph{Spitzer}/GLIMPSE dataset,
Gao, Jiang \& Li (2009) found that the mid-IR extinction
shows variation with \emph{Galactic longitude} which appears to
correlate with the locations of spiral arms.
Zasowski et al.\ (2009) also found that
the mid-IR extinction varies with \emph{longitude}.
The present work has not yet studied
the longitudinal variation of the mid-IR extinction.
However, this work seems to suggest no noticeable
\emph{latitudinal} variation of the mid-IR extinction
as revealed by the general consistency
between that determined
from the \emph{Spitzer}/GLIMPSE dataset (Galactic plane)
and that from the \emph{WISE} determination (all-sky).

\subsection{Comparison with interstellar dust models}

In Figure~\ref{extcurve}, the extinction curves calculated by three dust models are plotted, two of which are from the classical dust model by WD01 with $\RV=3.1$ and $\RV=5.5$ respectively, and the other is the newly developed dust model with an addition of micrometer-sized ice grains by \cite{WangLi15a}.

Our result matches pretty well with the extinction curve of the WD01 model for $\RV=5.5$ in most of the wavebands. The extinction is slightly lower than the model at $\lambda < 7.0\mum$, closely consistent in [8.0], S9W, W3 and [24]. But the discrepancy in the W4 band is significant. Meanwhile, our result in the W4 band matches the \cite{WangLi15a} model reasonably well, even better than the WD01 $\RV=5.5$ model if we consider the \emph{WISE}/W4 band at 22$\mum$.

The WD01 $R_V=5.5$ model assumes a mixture of
amorphous silicate dust and carbonaceous dust.\footnote{%
   The carbonaceous grain population was assumed to
   extend from grains with graphitic properties at radii
   $a>0.01\mum$, down to particles with PAH-like
   properties at very small sizes (see \cite{LiDraine2001}).
   }
Compared to the WD01 $R_V=3.1$ model
for the Galactic average diffuse ISM,
the $R_V=5.5$ model results in a much flatter
UV/optical extinction curve and is more representative
of dense clouds.
If the flat mid-IR extinction is indeed
``universally'' valid for both diffuse and dense regions,
the ice grain model of \cite{WangLi15a}
would be more favorable
as it reproduces the steep UV/optical extinction of
the diffuse ISM as well as the flat mid-IR extinction.
This model assumes a mixture of amorphous silicate grains,
carbonaceous grains and $\mu$m-sized ice grains.
According to this model, a population of spherical
ice grains with typical radii of $\simali$4$\mum$
and an oxygen abundance of O/H\,=\,160$\ppm$
are capable of accounting for
the observed flat extinction
at $\simali$3--8$\mum$
and produce no noticeable absorption feature
at $\simali$3.1$\mum$.
The 3.1$\mum$ absorption feature,
characteristic of sub-$\mu$m-sized H$_2$O ice grains,
is ubiquitously seen in dense molecular clouds.
This feature is not seen in the Galactic diffuse ISM
since sub-$\mu$m-sized H$_2$O ice grains are not expected
to be present in the diffuse ISM: they are easily
destroyed by photosputtering in the diffuse ISM
in a time scale ($\taupd$) of several hundred years
(e.g., see \cite{Westley1995}, \cite{Oberg2009}).
For $\mu$m-sized ice grains, the 3.1$\mum$ absorption feature
is much weaker and the photosputtering lifetime is much longer.
Wang et al.\ (2015a) argued that,
with a photosputtering lifetime of
$\taupd$\,$\approx$\,$5.8\times10^6$--$2.9\times10^7\yr$
longer than or comparable to
the turnover timescale of molecular clouds
of $\taumc$\,$\approx$\,$3\times10^6$--$2\times10^7\yr$
implied by the observed large depletions
of Si and Fe elements in the diffuse ISM
(see \cite{Draine1990}),
H$_2$O ice grains of $a\gtsim4\mum$
could be present in the diffuse ISM
through rapid exchange of material
between dense molecular clouds
where they form
and diffuse clouds
where they are destroyed by
photosputtering.

As shown in Figure~\ref{extcurve},
the extinction curve predicted from the ice grain model
exhibits a narrow, minor structure at $\simali$2.8$\mum$
arising from the scattering of
the O--H stretch of $\mu$m-sized H$_2$O ice.
The extinction toward the Galactic center
derived by Lutz (1999) based on
the recombination lines of atomic H
was not sufficiently resolved in wavelength
to rule out this structure.\footnote{%
   \cite{Whittet1997} reported
   the detection of a shallow feature
   centered at $\simali$2.75$\mum$
   in the near-IR spectrum
   of Cygnus OB2 No.\,12 obtained by
   the \emph{Kuiper Airborne Observatory} (KAO)
   and \emph{ISO}/SWS.
   They tentatively attributed this feature
   to the OH groups of hydrated silicates.
   However, \cite{Whittet2001} found that
   this structure was an artifact
   caused by the calibration uncertainty.
   }
Future high resolution spectroscopic search
for the $\simali$2.8$\mum$ feature in lines
of sight toward diffuse regions would be of
great value in testing the ice grain model
of \cite{WangLi15a}.
Finally, we also note that the observed flat
mid-IR extinction could also be reproduced by
a population of very large, $\mu$m-sized graphitic grains
with a typical size of $\simali$1.2$\mum$
\citep{WangLi15b}.
The mid-IR extinction predicted from
the $\mu$m-sized graphite model of \cite{WangLi15b}
closely resembles that of the WD01 $R_V=5.5$ model,
with the 22$\mum$ extinction appreciably lower than
the \emph{WISE}/W4 band detection.
The advantage of the ice grain model of
\cite{WangLi15a} is that, it not only better fits
the \emph{WISE}/W4 band extinction at 22$\mum$,
but also could serve as a reservoir for
the missing O atoms.\footnote{%
   The ISM seems to have a significant
   \emph{surplus} of oxygen \citep{Jenkins2009}:
   independent of the adopted interstellar
   reference abundance,
   the total number of O atoms
   depleted from the gas phase
   far exceeds that tied up in
   silicates and metallic oxides,
   the major solid-phase reservoirs of O,
   by as much as $\simali$160$\ppm$ of O/H
   \citep{Whittet2010}.
   }

\section{Summary}

The average mid-infrared extinction law is derived
with high precision in the four \emph{WISE} bands,
the four \emph{Spitzer}/IRAC bands,
the \emph{Spitzer}/MIPS 24$\mum$ band
and the\emph{AKARI}/S9W band
for the whole sky
using G- and K-type giants
as the extinction tracer.
This derivation is based on
the color-excess method,
with the intrinsic stellar color indices
determined from the stellar photospheric spectra
obtained by APOGEE.

The major results of this paper are as follows:
\begin{enumerate}
\item The stellar intrinsic colors
      $C^0_{\rm {\Ks\lambda}}\equiv (\lambda-\Ks)_{\rm intrinsic}$
      are fitted as simple analytical functions of
      the stellar effective temperature $\Teff$
      in the $\sim$3600--5200\,K temperature range
      for a given near- and mid-IR photometric band
      of an effective wavelength $\lambda$.
%
\item We derive the 2MASS near-IR color-excess ratio to be
      $E_{\rm JH}/E_{\rm J\Ks}\approx0.652$,
      corresponding to a power exponent of
      $\alpha\approx1.79$
      for a power-law near-IR extinction of
      $A_\lambda \propto \lambda^{-\alpha}$.
%
\item We derive the mean extinction
      (relative to the K$_{\rm S}$ band) to be
      ${A_{\rm {W1}}}/\AKs\approx0.591$,
      ${A_{\rm {W2}}}/\AKs\approx0.463$,
      ${A_{\rm {W3}}}/\AKs\approx0.537$,
      ${A_{\rm {W4}}}/\AKs\approx0.364$,
      ${A_{\rm {[3.6]}}}/\AKs\approx0.553$,
      ${A_{\rm {[4.5]}}}/\AKs\approx0.461$,
      ${A_{\rm {[5.8]}}}/\AKs\approx0.389$,
      ${A_{\rm {[8.0]}}}/\AKs\approx0.426$,
      ${A_{\rm {[24]}}}/\AKs\approx0.264$
      and ${A_{\rm S9W}}/\AKs\approx0.530$ respectively
      for the four \emph{WISE} bands
      at 3.4, 4.6, 12 and 22$\mum$,
      the four \emph{Spitzer}/IRAC bands
      at 3.6, 4.5, 5.8 and 8$\mum$,
      the \emph{Spitzer}/MIPS24 band at 23.7$\mum$,
      and for the first time,
      the \emph{AKARI}/S9W band at 8.23$\mum$.
\item In agreement with previous works,
      the mid-IR extinction law derived here
      is flat in the $\simali$3--8$\mum$ wavelength range
      and closely resembles the WD01 model extinction
      curve of $\RV=5.5$ and the ice grain model of
      \cite{WangLi15a}, with the former somewhat
      underpredicting the \emph{WISE}/W4 band extinction
      at 22$\mum$ while the latter closely reproducing
      the observed extinction in the entire wavelength
      range of $\simali$3--25$\mum$
      and therefore the ice grain model appears
      more favorable.
      The 8.23$\mum$ AKARI/S9W extinction
      and the 11.56$\mum$ \emph{WISE}/W3 extinction appear
      consistent with the absorption profile of the
      \cite{DraineLee1984} astronomical silicates.
\item  Although some previous works have shown that
       the mid-IR extinction law seems to vary with
       the extinction depth $\AKs$,
       no noticeable variation has been found in this work.
       It has previously also been shown that the mid-IR
       extinction law appears to vary
       with \emph{Galactic longitude}.
       The present work has not yet studied
       the longitudinal variation.
       However, this work seems to suggest no noticeable variation
       of the mid-IR extinction with \emph{latitude}
       as revealed by the general consistency
       between that determined
       from the \emph{Spitzer}/GLIMPSE dataset (Galactic plane)
       and that from the \emph{WISE} determination (all-sky).
\end{enumerate}


\acknowledgments{We thank Drs. Shuang~Gao, Yang~Huang, Mingjie~Jian,
                 Xiaowei~Liu, He~Zhao and the anonymous
                 referee for their very helpful suggestions/comments.
                 This work is supported by NSFC through Projects
                 11173007, 11373015, 11533002,
                 and 973 Program 2014CB845702.
                 AL and SW are supported in part by
                 NSF AST-1109039, and NNX13AE63G.
}

\emph{Facilities:} \facility{\emph{AKARI}}, \facility{\emph{Spitzer}(IRAC)}, \facility{\emph{Spitzer}(MIPS)}, \facility{\emph{WISE}}, \facility{2MASS}.

\clearpage

\bibliographystyle{aasjournal}
\bibliography{bib}

\begin{thebibliography}{}
\expandafter\ifx\csname natexlab\endcsname\relax\def\natexlab#1{#1}\fi

\bibitem[{{Alam} {et~al.}(2015){Alam}, {Albareti}, {Allende Prieto}, {Anders},
  {Anderson}, {Anderton}, {Andrews}, {Armengaud}, {Aubourg}, {Bailey}, \&
  et~al.}]{2015SDSSdr12}
{Alam}, S., {Albareti}, F.~D., {Allende Prieto}, C., {et~al.} 2015, \apjs, 219,
  12

\bibitem[{{Ascenso} {et~al.}(2013){Ascenso}, {Lada}, {Alves},
  {Rom{\'a}n-Z{\'u}{\~n}iga}, \& {Lombardi}}]{Ascenso2013}
{Ascenso}, J., {Lada}, C.~J., {Alves}, J., {Rom{\'a}n-Z{\'u}{\~n}iga}, C.~G.,
  \& {Lombardi}, M. 2013, \aap, 549, A135

\bibitem[{{Bernat}(1981)}]{Bernat1981}
{Bernat}, A.~P. 1981, \apj, 246, 184

\bibitem[{{Bessell} \& {Brett}(1988)}]{Bessell88}
{Bessell}, M.~S., \& {Brett}, J.~M. 1988, \pasp, 100, 1134

\bibitem[{{Boogert} {et~al.}(2013){Boogert}, {Chiar}, {Knez}, {{\"O}berg},
  {Mundy}, {Pendleton}, {Tielens}, \& {van Dishoeck}}]{Boogert2013}
{Boogert}, A.~C.~A., {Chiar}, J.~E., {Knez}, C., {et~al.} 2013, \apj, 777, 73

\bibitem[{{Boogert} {et~al.}(2015){Boogert}, {Gerakines}, \&
  {Whittet}}]{Boogert2015}
{Boogert}, A.~C.~A., {Gerakines}, P.~A., \& {Whittet}, D.~C.~B. 2015, \araa,
  53, 541

\bibitem[{{Boogert} {et~al.}(2011){Boogert}, {Huard}, {Cook}, {Chiar}, {Knez},
  {Decin}, {Blake}, {Tielens}, \& {van Dishoeck}}]{Boogert2011}
{Boogert}, A.~C.~A., {Huard}, T.~L., {Cook}, A.~M., {et~al.} 2011, \apj, 729,
  92

\bibitem[{{Cardelli} {et~al.}(1989){Cardelli}, {Clayton}, \&
  {Mathis}}]{Cardelli89}
{Cardelli}, J.~A., {Clayton}, G.~C., \& {Mathis}, J.~S. 1989, \apj, 345, 245

\bibitem[{{Carey} {et~al.}(2009){Carey}, {Noriega-Crespo}, {Mizuno}, {Shenoy},
  {Paladini}, {Kraemer}, {Price}, {Flagey}, {Ryan}, {Ingalls}, {Kuchar},
  {Pinheiro Gon{\c c}alves}, {Indebetouw}, {Billot}, {Marleau}, {Padgett},
  {Rebull}, {Bressert}, {Ali}, {Molinari}, {Martin}, {Berriman}, {Boulanger},
  {Latter}, {Miville-Deschenes}, {Shipman}, \& {Testi}}]{Carey2009}
{Carey}, S.~J., {Noriega-Crespo}, A., {Mizuno}, D.~R., {et~al.} 2009, \pasp,
  121, 76

\bibitem[{{Chapman} {et~al.}(2009){Chapman}, {Mundy}, {Lai}, \&
  {Evans}}]{Chapman2009}
{Chapman}, N.~L., {Mundy}, L.~G., {Lai}, S.-P., \& {Evans}, II, N.~J. 2009,
  \apj, 690, 496

\bibitem[{{Chen} {et~al.}(2014){Chen}, {Liu}, {Yuan}, {Zhang}, {Schultheis},
  {Jiang}, {Huang}, {Xiang}, {Zhao}, {Yao}, \& {Lu}}]{ChenBQ2014}
{Chen}, B.-Q., {Liu}, X.-W., {Yuan}, H.-B., {et~al.} 2014, \mnras, 443, 1192

\bibitem[{{Chiar} \& {Tielens}(2006)}]{Chiar2006}
{Chiar}, J.~E., \& {Tielens}, A.~G.~G.~M. 2006, \apj, 637, 774

\bibitem[{{Chiar} {et~al.}(2007){Chiar}, {Ennico}, {Pendleton}, {Boogert},
  {Greene}, {Knez}, {Lada}, {Roellig}, {Tielens}, {Werner}, \&
  {Whittet}}]{Chiar2007}
{Chiar}, J.~E., {Ennico}, K., {Pendleton}, Y.~J., {et~al.} 2007, \apjl, 666,
  L73

\bibitem[{{Chiar} {et~al.}(2011){Chiar}, {Pendleton}, {Allamandola}, {Boogert},
  {Ennico}, {Greene}, {Geballe}, {Keane}, {Lada}, {Mason}, {Roellig},
  {Sandford}, {Tielens}, {Werner}, {Whittet}, {Decin}, \&
  {Eriksson}}]{Chiar2011}
{Chiar}, J.~E., {Pendleton}, Y.~J., {Allamandola}, L.~J., {et~al.} 2011, \apj,
  731, 9

\bibitem[{{Churchwell} {et~al.}(2009){Churchwell}, {Babler}, {Meade},
  {Whitney}, {Benjamin}, {Indebetouw}, {Cyganowski}, {Robitaille}, {Povich},
  {Watson}, \& {Bracker}}]{Churchwell2009}
{Churchwell}, E., {Babler}, B.~L., {Meade}, M.~R., {et~al.} 2009, \pasp, 121,
  213

\bibitem[{{Cohen} {et~al.}(2003){Cohen}, {Wheaton}, \& {Megeath}}]{Cohen2003}
{Cohen}, M., {Wheaton}, W.~A., \& {Megeath}, S.~T. 2003, \aj, 126, 1090

\bibitem[{{Cutri} {et~al.}(2003){Cutri}, {Skrutskie}, {van Dyk}, {Beichman},
  {Carpenter}, {Chester}, {Cambresy}, {Evans}, {Fowler}, {Gizis}, {Howard},
  {Huchra}, {Jarrett}, {Kopan}, {Kirkpatrick}, {Light}, {Marsh}, {McCallon},
  {Schneider}, {Stiening}, {Sykes}, {Weinberg}, {Wheaton}, {Wheelock}, \&
  {Zacarias}}]{Cutri2003}
{Cutri}, R.~M., {Skrutskie}, M.~F., {van Dyk}, S., {et~al.} 2003, {2MASS All
  Sky Catalog of point sources.}

\bibitem[{{Cutri} {et~al.}(2013){Cutri}, {Wright}, {Conrow}, {Fowler},
  {Eisenhardt}, {Grillmair}, {Kirkpatrick}, {Masci}, {McCallon}, {Wheelock},
  {Fajardo-Acosta}, {Yan}, {Benford}, {Harbut}, {Jarrett}, {Lake}, {Leisawitz},
  {Ressler}, {Stanford}, {Tsai}, {Liu}, {Helou}, {Mainzer}, {Gettings},
  {Gonzalez}, {Hoffman}, {Marsh}, {Padgett}, {Skrutskie}, {Beck}, {Papin}, \&
  {Wittman}}]{2013wise}
{Cutri}, R.~M., {Wright}, E.~L., {Conrow}, T., {et~al.} 2013, {Explanatory
  Supplement to the AllWISE Data Release Products}, Tech. rep.

\bibitem[{{Davenport} {et~al.}(2014){Davenport}, {Ivezi{\'c}}, {Becker},
  {Ruan}, {Hunt-Walker}, {Covey}, {Lewis}, {AlSayyad}, \&
  {Anderson}}]{Davenport2014}
{Davenport}, J.~R.~A., {Ivezi{\'c}}, {\v Z}., {Becker}, A.~C., {et~al.} 2014,
  \mnras, 440, 3430

\bibitem[{{Draine}(1990)}]{Draine1990}
{Draine}, B.~T. 1990, in Astronomical Society of the Pacific Conference Series,
  Vol.~12, The Evolution of the Interstellar Medium, ed. L.~{Blitz}, 193--205

\bibitem[{{Draine} \& {Lee}(1984)}]{DraineLee1984}
{Draine}, B.~T., \& {Lee}, H.~M. 1984, \apj, 285, 89

\bibitem[{{Ducati} {et~al.}(2001){Ducati}, {Bevilacqua}, {Rembold}, \&
  {Ribeiro}}]{Ducati2001}
{Ducati}, J.~R., {Bevilacqua}, C.~M., {Rembold}, S.~B., \& {Ribeiro}, D. 2001,
  \apj, 558, 309

\bibitem[{Efron(1979)}]{Efron79}
Efron, B. 1979, Annals of Statistics, 7, 1

\bibitem[{{Eisenstein} {et~al.}(2011){Eisenstein}, {Weinberg}, {Agol},
  {Aihara}, {Allende Prieto}, {Anderson}, {Arns}, {Aubourg}, {Bailey},
  {Balbinot}, \& et~al.}]{Eisenstein2011}
{Eisenstein}, D.~J., {Weinberg}, D.~H., {Agol}, E., {et~al.} 2011, \aj, 142, 72

\bibitem[{{Flaherty} {et~al.}(2007){Flaherty}, {Pipher}, {Megeath}, {Winston},
  {Gutermuth}, {Muzerolle}, {Allen}, \& {Fazio}}]{Flaherty07}
{Flaherty}, K.~M., {Pipher}, J.~L., {Megeath}, S.~T., {et~al.} 2007, \apj, 663,
  1069

\bibitem[{{Gao} {et~al.}(2009){Gao}, {Jiang}, \& {Li}}]{Gao09}
{Gao}, J., {Jiang}, B.~W., \& {Li}, A. 2009, \apj, 707, 89

\bibitem[{{Gao} {et~al.}(2013){Gao}, {Jiang}, {Li}, \& {Xue}}]{gao2013}
{Gao}, J., {Jiang}, B.~W., {Li}, A., \& {Xue}, M.~Y. 2013, \apj, 776, 7

\bibitem[{{Gunn} {et~al.}(2006){Gunn}, {Siegmund}, {Mannery}, {Owen}, {Hull},
  {Leger}, {Carey}, {Knapp}, {York}, {Boroski}, {Kent}, {Lupton}, {Rockosi},
  {Evans}, {Waddell}, {Anderson}, {Annis}, {Barentine}, {Bartoszek}, {Bastian},
  {Bracker}, {Brewington}, {Briegel}, {Brinkmann}, {Brown}, {Carr},
  {Czarapata}, {Drennan}, {Dombeck}, {Federwitz}, {Gillespie}, {Gonzales},
  {Hansen}, {Harvanek}, {Hayes}, {Jordan}, {Kinney}, {Klaene}, {Kleinman},
  {Kron}, {Kresinski}, {Lee}, {Limmongkol}, {Lindenmeyer}, {Long}, {Loomis},
  {McGehee}, {Mantsch}, {Neilsen}, {Neswold}, {Newman}, {Nitta}, {Peoples},
  {Pier}, {Prieto}, {Prosapio}, {Rivetta}, {Schneider}, {Snedden}, \&
  {Wang}}]{Gunn2006}
{Gunn}, J.~E., {Siegmund}, W.~A., {Mannery}, E.~J., {et~al.} 2006, \aj, 131,
  2332

\bibitem[{{Gutermuth} \& {Heyer}(2015)}]{Gutermuth2015}
{Gutermuth}, R.~A., \& {Heyer}, M. 2015, \aj, 149, 64

\bibitem[{{Hora} {et~al.}(2008){Hora}, {Carey}, {Surace}, {Marengo},
  {Lowrance}, {Glaccum}, {Lacy}, {Reach}, {Hoffmann}, {Barmby}, {Willner},
  {Fazio}, {Megeath}, {Allen}, {Bhattacharya}, \& {Quijada}}]{Hora2008}
{Hora}, J.~L., {Carey}, S., {Surace}, J., {et~al.} 2008, \pasp, 120, 1233

\bibitem[{{Indebetouw} {et~al.}(2005){Indebetouw}, {Mathis}, {Babler}, {Meade},
  {Watson}, {Whitney}, {Wolff}, {Wolfire}, {Cohen}, {Bania}, {Benjamin},
  {Clemens}, {Dickey}, {Jackson}, {Kobulnicky}, {Marston}, {Mercer},
  {Stauffer}, {Stolovy}, \& {Churchwell}}]{Indebetouw05}
{Indebetouw}, R., {Mathis}, J.~S., {Babler}, B.~L., {et~al.} 2005, \apj, 619,
  931

\bibitem[{{Ishihara} {et~al.}(2010){Ishihara}, {Onaka}, {Kataza}, {Salama},
  {Alfageme}, {Cassatella}, {Cox}, {Garc{\'{\i}}a-Lario}, {Stephenson},
  {Cohen}, {Fujishiro}, {Fujiwara}, {Hasegawa}, {Ita}, {Kim}, {Matsuhara},
  {Murakami}, {M{\"u}ller}, {Nakagawa}, {Ohyama}, {Oyabu}, {Pyo}, {Sakon},
  {Shibai}, {Takita}, {Tanab{\'e}}, {Uemizu}, {Ueno}, {Usui}, {Wada},
  {Watarai}, {Yamamura}, \& {Yamauchi}}]{Ishihara2010}
{Ishihara}, D., {Onaka}, T., {Kataza}, H., {et~al.} 2010, \aap, 514, A1

\bibitem[{{Jenkins}(2009)}]{Jenkins2009}
{Jenkins}, E.~B. 2009, \apj, 700, 1299

\bibitem[{{Jiang} {et~al.}(2003){Jiang}, {Omont}, {Ganesh}, {Simon}, \&
  {Schuller}}]{Jiang03}
{Jiang}, B.~W., {Omont}, A., {Ganesh}, S., {Simon}, G., \& {Schuller}, F. 2003,
  \aap, 400, 903

\bibitem[{{Li} \& {Draine}(2001)}]{LiDraine2001}
{Li}, A., \& {Draine}, B.~T. 2001, \apj, 554, 778

\bibitem[{{Lutz}(1999)}]{Lutz99}
{Lutz}, D. 1999, in ESA Special Publication, Vol. 427, The Universe as Seen by
  ISO, ed. P.~{Cox} \& M.~{Kessler}, 623

\bibitem[{{Lutz} {et~al.}(1996){Lutz}, {Feuchtgruber}, {Genzel}, {Kunze},
  {Rigopoulou}, {Spoon}, {Wright}, {Egami}, {Katterloher}, {Sturm},
  {Wieprecht}, {Sternberg}, {Moorwood}, \& {de Graauw}}]{Lutz96}
{Lutz}, D., {Feuchtgruber}, H., {Genzel}, R., {et~al.} 1996, \aap, 315, L269

\bibitem[{{Majewski} {et~al.}(2011){Majewski}, {Zasowski}, \&
  {Nidever}}]{2011RJCE}
{Majewski}, S.~R., {Zasowski}, G., \& {Nidever}, D.~L. 2011, \apj, 739, 25

\bibitem[{{McClure}(2009)}]{McClure2009}
{McClure}, M. 2009, \apjl, 693, L81

\bibitem[{{Murakami} {et~al.}(2007){Murakami}, {Baba}, {Barthel}, {Clements},
  {Cohen}, {Doi}, {Enya}, {Figueredo}, {Fujishiro}, {Fujiwara}, {Fujiwara},
  {Garcia-Lario}, {Goto}, {Hasegawa}, {Hibi}, {Hirao}, {Hiromoto}, {Hong},
  {Imai}, {Ishigaki}, {Ishiguro}, {Ishihara}, {Ita}, {Jeong}, {Jeong},
  {Kaneda}, {Kataza}, {Kawada}, {Kawai}, {Kawamura}, {Kessler}, {Kester},
  {Kii}, {Kim}, {Kim}, {Kobayashi}, {Koo}, {Kwon}, {Lee}, {Lorente}, {Makiuti},
  {Matsuhara}, {Matsumoto}, {Matsuo}, {Matsuura}, {M{\"u}ller}, {Murakami},
  {Nagata}, {Nakagawa}, {Naoi}, {Narita}, {Noda}, {Oh}, {Ohnishi}, {Ohyama},
  {Okada}, {Okuda}, {Oliver}, {Onaka}, {Ootsubo}, {Oyabu}, {Pak}, {Park},
  {Pearson}, {Rowan-Robinson}, {Saito}, {Sakon}, {Salama}, {Sato}, {Savage},
  {Serjeant}, {Shibai}, {Shirahata}, {Sohn}, {Suzuki}, {Takagi}, {Takahashi},
  {Tanab{\'e}}, {Takeuchi}, {Takita}, {Thomson}, {Uemizu}, {Ueno}, {Usui},
  {Verdugo}, {Wada}, {Wang}, {Watabe}, {Watarai}, {White}, {Yamamura},
  {Yamauchi}, \& {Yasuda}}]{Murakami2007}
{Murakami}, H., {Baba}, H., {Barthel}, P., {et~al.} 2007, \pasj, 59, 369

\bibitem[{{{\"O}berg} {et~al.}(2009){{\"O}berg}, {Linnartz}, {Visser}, \& {van
  Dishoeck}}]{Oberg2009}
{{\"O}berg}, K.~I., {Linnartz}, H., {Visser}, R., \& {van Dishoeck}, E.~F.
  2009, \apj, 693, 1209

\bibitem[{{Onaka} {et~al.}(2007){Onaka}, {Matsuhara}, {Wada}, {Fujishiro},
  {Fujiwara}, {Ishigaki}, {Ishihara}, {Ita}, {Kataza}, {Kim}, {Matsumoto},
  {Murakami}, {Ohyama}, {Oyabu}, {Sakon}, {Tanab{\'e}}, {Takagi}, {Uemizu},
  {Ueno}, {Usui}, {Watarai}, {Cohen}, {Enya}, {Ootsubo}, {Pearson}, {Takeyama},
  {Yamamuro}, \& {Ikeda}}]{Onaka2007}
{Onaka}, T., {Matsuhara}, H., {Wada}, T., {et~al.} 2007, \pasj, 59, 401

\bibitem[{{Ram{\'{\i}}rez} \& {Mel{\'e}ndez}(2005)}]{Ramirez05}
{Ram{\'{\i}}rez}, I., \& {Mel{\'e}ndez}, J. 2005, \apj, 626, 465

\bibitem[{{Rieke} \& {Lebofsky}(1985)}]{Rieke85}
{Rieke}, G.~H., \& {Lebofsky}, M.~J. 1985, \apj, 288, 618

\bibitem[{{Rom{\'a}n-Z{\'u}{\~n}iga} {et~al.}(2007){Rom{\'a}n-Z{\'u}{\~n}iga},
  {Lada}, {Muench}, \& {Alves}}]{Roman-Zuniga2007}
{Rom{\'a}n-Z{\'u}{\~n}iga}, C.~G., {Lada}, C.~J., {Muench}, A., \& {Alves},
  J.~F. 2007, \apj, 664, 357

\bibitem[{{Schlegel} {et~al.}(1998){Schlegel}, {Finkbeiner}, \&
  {Davis}}]{Schlegel98}
{Schlegel}, D.~J., {Finkbeiner}, D.~P., \& {Davis}, M. 1998, \apj, 500, 525

\bibitem[{{Wang} {et~al.}(2013){Wang}, {Gao}, {Jiang}, {Li}, \&
  {Chen}}]{Wang2013}
{Wang}, S., {Gao}, J., {Jiang}, B.~W., {Li}, A., \& {Chen}, Y. 2013, \apj, 773,
  30

\bibitem[{{Wang} \& {Jiang}(2014)}]{WJ14}
{Wang}, S., \& {Jiang}, B.~W. 2014, \apjl, 788, L12

\bibitem[{{Wang} {et~al.}(2014){Wang}, {Li}, \& {Jiang}}]{WLJ14}
{Wang}, S., {Li}, A., \& {Jiang}, B.~W. 2014, \planss, 100, 32

\bibitem[{{Wang} {et~al.}(2015{\natexlab{a}}){Wang}, {Li}, \&
  {Jiang}}]{WangLi15a}
---. 2015{\natexlab{a}}, \mnras, 454, 569

\bibitem[{{Wang} {et~al.}(2015{\natexlab{b}}){Wang}, {Li}, \&
  {Jiang}}]{WangLi15b}
---. 2015{\natexlab{b}}, \apj, 811, 38

\bibitem[{{Weingartner} \& {Draine}(2001)}]{WD01}
{Weingartner}, J.~C., \& {Draine}, B.~T. 2001, \apj, 548, 296

\bibitem[{{Westley} {et~al.}(1995){Westley}, {Baragiola}, {Johnson}, \&
  {Baratta}}]{Westley1995}
{Westley}, M.~S., {Baragiola}, R.~A., {Johnson}, R.~E., \& {Baratta}, G.~A.
  1995, \nat, 373, 405

\bibitem[{{Whittet}(2010)}]{Whittet2010}
{Whittet}, D.~C.~B. 2010, \apj, 710, 1009

\bibitem[{{Whittet} {et~al.}(2001){Whittet}, {Pendleton}, {Gibb}, {Boogert},
  {Chiar}, \& {Nummelin}}]{Whittet2001}
{Whittet}, D.~C.~B., {Pendleton}, Y.~J., {Gibb}, E.~L., {et~al.} 2001, \apj,
  550, 793

\bibitem[{{Whittet} {et~al.}(1997){Whittet}, {Boogert}, {Gerakines}, {Schutte},
  {Tielens}, {de Graauw}, {Prusti}, {van Dishoeck}, {Wesselius}, \&
  {Wright}}]{Whittet1997}
{Whittet}, D.~C.~B., {Boogert}, A.~C.~A., {Gerakines}, P.~A., {et~al.} 1997,
  \apj, 490, 729

\bibitem[{{Wright} {et~al.}(2010){Wright}, {Eisenhardt}, {Mainzer}, {Ressler},
  {Cutri}, {Jarrett}, {Kirkpatrick}, {Padgett}, {McMillan}, {Skrutskie},
  {Stanford}, {Cohen}, {Walker}, {Mather}, {Leisawitz}, {Gautier}, {McLean},
  {Benford}, {Lonsdale}, {Blain}, {Mendez}, {Irace}, {Duval}, {Liu}, {Royer},
  {Heinrichsen}, {Howard}, {Shannon}, {Kendall}, {Walsh}, {Larsen}, {Cardon},
  {Schick}, {Schwalm}, {Abid}, {Fabinsky}, {Naes}, \& {Tsai}}]{Wright2010}
{Wright}, E.~L., {Eisenhardt}, P.~R.~M., {Mainzer}, A.~K., {et~al.} 2010, \aj,
  140, 1868

\bibitem[{{Yuan} {et~al.}(2013){Yuan}, {Liu}, \& {Xiang}}]{Yuan2013}
{Yuan}, H.~B., {Liu}, X.~W., \& {Xiang}, M.~S. 2013, \mnras, 430, 2188

\bibitem[{{Zasowski} {et~al.}(2009){Zasowski}, {Majewski}, {Indebetouw},
  {Meade}, {Nidever}, {Patterson}, {Babler}, {Skrutskie}, {Watson}, {Whitney},
  \& {Churchwell}}]{Zasowski09}
{Zasowski}, G., {Majewski}, S.~R., {Indebetouw}, R., {et~al.} 2009, \apj, 707,
  510

\end{thebibliography}

\begin{landscape}
\begin{table}
\caption{\label{data quality}The photometric catalogs in use and their cross-identifications with the APOGEE catalog.}
\begin{tabular}{m{64pt}<{\centering}|cc|cccc|cccc|c}
\hline
\hline
Survey                       &  \multicolumn{2}{c|}{AKARI}   & \multicolumn{4}{c|}{\emph{WISE}}   &  \multicolumn{4}{c|}{\emph{Spitzer}/GLIMPSE}  & {\emph{Spitzer}/MIPSGAL} \\
\hline
Bands                        & S9W & L18W &  W1  &  W2  &  W3  &  W4    & [3.6] & [4.5] & [5.8] & [8.0]  & [24] \\
\hline
$\lambda_{\rm eff}$ ($\mum$)  & 8.23 & 17.61 & 3.35 & 4.60 & 11.56 & 22.09 & 3.55 & 4.49 & 5.73 & 7.87 & 23.68 \\
\hline
 area &  \multicolumn{2}{c|}{All sky} & \multicolumn{4}{c|}{All sky} & \multicolumn{4}{c|}{$-105\degr<l<65\degr$} & $-68\degr<l<69\degr$  \\
 &  \multicolumn{2}{c|}{ } & \multicolumn{4}{c|}{ } & \multicolumn{4}{c|}{$|b|<5\degr$} &  $|b|<3\degr$  \\
\hline
$5\sigma$ limit (mag)        & 7.6 & 5.0 & 16.9 & 16.0 & 11.5 & 8.0   & 15.0 & 14.5 & 12.5 & 12.5 &       7.9\\
\hline
Cross radius                & \multicolumn{2}{c|}{3\arcsec} & \multicolumn{4}{c|}{1\arcsec} & \multicolumn{4}{c|}{\tablenotemark{\ast}} &  3\arcsec \\
\hline
No. of sources $\otimes$ APOGEE  & 4296 & 901 & 154842 & 154845 & 154735 & 154793 & 15058 & 15307 & 15251 & 15071 & 3045 \\
\hline
$\sigma_\lambda$ (mag)  & 0.2 & 0.3 & \multicolumn{4}{c|}{0.1} & \multicolumn{4}{c|}{0.1} & 0.2 \\
\hline
No. of sources qualified  & 1024 & 108 & 61734 & 61935 & 41510 & 2008 & 5411 & 5540 & 5474 & 5502 & 806 \\
\hline
\end{tabular}
\tablenotetext{\ast}{As the APOGEE catalog includes the Spitzer/IRAC photometry, we made no new cross-identification.}
\end{table}
\end{landscape}

\begin{landscape}
\begin{table}
\begin{center}
\caption{\label{Teffcompare}Comparison of $\CJKs^{0}$ of this work with that of \cite{Bessell88}(BB88) and WJ14}
\begin{tabular}{lccccccccccc}
\tableline \tableline
    $\Teff(\K)$        & 3630 & 3710 & 3780 & 3820 & 3980 & 4080 & 4320 & 4500 & 4610 & 4810 & 4960 \\
\tableline
    This work          & 1.21 & 1.15 & 1.10 & 1.07 & 0.97 & 0.91 & 0.78 & 0.70 & 0.65 & 0.57 & 0.52 \\
\tableline
    BB88 & 1.13 & 1.08 & 1.05 & 1.01 & 0.95 & 0.88 & 0.82 & 0.74 & 0.68 & 0.63 & 0.58 \\
\tableline
    WJ14 & 1.43 & 1.35 & 1.28 & 1.25 & 1.11 & 1.03 & 0.87 & 0.78 & 0.73 & 0.66 & 0.62 \\
\tableline
\end{tabular}
\end{center}
\end{table}
\end{landscape}

\begin{landscape}
\begin{table}
\caption{\label{tab_colors}The intrinsic color indices at typical $\Teff$}
    \begin{tabular}{ccccccccccccc}
    \hline
    \hline
          & $C^0_{\rm{J\Ks}}$  & $C^0_{\rm{JH}}$ & $C^0_{\rm{\Ks W1}}$ & $C^0_{\rm{\Ks W2}}$ & $C^0_{\rm{\Ks W3}}$ & $C^0_{\rm{\Ks W4}}$ & $C^0_{\rm{\Ks S9W}}$ & $C^0_{\rm{\Ks [3.6]}}$ & $C^0_{\rm{\Ks [4.5]}}$ & $C^0_{\rm{\Ks [5.8]}}$ & $C^0_{\rm{\Ks [8.0]}}$ & $C^0_{\rm{\Ks [24]}}$ \\
    \hline
    3600K  & 1.254  & 0.961  & 0.131  & -0.052  & 0.184  & 0.347  & 0.331  & 0.145  & -0.122  & 0.113  & 0.130  & 0.126  \\
    3700K  & 1.172  & 0.911  & 0.124  & -0.051  & 0.168  & 0.318  & 0.308  & 0.135  & -0.109  & 0.103  & 0.117  & 0.118  \\
    3800K  & 1.096  & 0.864  & 0.117  & -0.049  & 0.154  & 0.292  & 0.287  & 0.124  & -0.097  & 0.094  & 0.106  & 0.110  \\
    3900K  & 1.026  & 0.820  & 0.110  & -0.047  & 0.142  & 0.267  & 0.267  & 0.114  & -0.087  & 0.084  & 0.096  & 0.103  \\
    4000K  & 0.961  & 0.777  & 0.104  & -0.045  & 0.131  & 0.245  & 0.249  & 0.105  & -0.078  & 0.075  & 0.088  & 0.095  \\
    4100K  & 0.902  & 0.737  & 0.099  & -0.042  & 0.122  & 0.225  & 0.233  & 0.096  & -0.070  & 0.067  & 0.080  & 0.087  \\
    4200K  & 0.847  & 0.700  & 0.094  & -0.039  & 0.114  & 0.207  & 0.219  & 0.087  & -0.064  & 0.058  & 0.074  & 0.079  \\
    4300K  & 0.796  & 0.664  & 0.089  & -0.036  & 0.108  & 0.191  & 0.206  & 0.078  & -0.060  & 0.050  & 0.069  & 0.072  \\
    4400K  & 0.750  & 0.630  & 0.085  & -0.032  & 0.103  & 0.177  & 0.195  & 0.070  & -0.057  & 0.042  & 0.066  & 0.064  \\
    4500K  & 0.707  & 0.598  & 0.082  & -0.028  & 0.100  & 0.165  & 0.186  & 0.062  & -0.055  & 0.035  & 0.064  & 0.056  \\
    4600K  & 0.668  & 0.567  & 0.079  & -0.024  & 0.099  & 0.155  & 0.179  & 0.055  & -0.055  & 0.028  & 0.062  & 0.048  \\
    4700K  & 0.631  & 0.538  & 0.076  & -0.019  & 0.099  & 0.148  & 0.173  & 0.047  & -0.057  & 0.021  & 0.063  & 0.041  \\
    4800K  & 0.598  & 0.511  & 0.075  & -0.014  & 0.101  & 0.142  & 0.169  & 0.040  & -0.060  & 0.015  & 0.064  & 0.033  \\
    4900K  & 0.567  & 0.485  & 0.073  & -0.008  & 0.104  & 0.139  & 0.167  & 0.034  & -0.065  & 0.009  & 0.067  & 0.025  \\
    5000K  & 0.539  & 0.460  & 0.072  & -0.002  & 0.109  & 0.137  & 0.167  & 0.028  & -0.071  & 0.003  & 0.071  & 0.017  \\
    5100K  & 0.513  & 0.437  & 0.072  & 0.004  & 0.116  & 0.138  & 0.168  & 0.022  & -0.078  & -0.002  & 0.076  & 0.009  \\
    5200K  & 0.489  & 0.415  & 0.072  & 0.011  & 0.124  & 0.141  & 0.171  & 0.016  & -0.087  & -0.007  & 0.082  & 0.002  \\
    \hline
    \end{tabular}%
\end{table}%
\end{landscape}

\begin{landscape}
\begin{table}
\begin{center}
\caption{The relative extinction in the mid-infrared bands and in comparison with other work}
\vspace{0.05in}
\begin{tabular}{lccccccc}
\tableline \tableline
                       & $E_{\Ks\lambda}/E_{J\Ks}$ & $A_{\lambda}/\AKs$ & $A_{\lambda}/\AKs$ & $A_{\lambda}/\AKs$ & $A_{\lambda}/\AKs$ & $A_{\lambda}/\AKs$ & $A_{\lambda}/\AKs$ \\
                       &  & ($A_{\rm J}/A_\Ks=2.72$) & ($A_{\rm J}/A_\Ks=2.52$) & (1)\tablenotemark{\ast} & (2)\tablenotemark{\ast} & (3)\tablenotemark{\ast} & (4)\tablenotemark{\ast} \\
\tableline
    \emph{WISE}/W1            &  0.238  &  0.591  &  0.638  &         &         &  0.63   &  0.60   \\
    \emph{WISE}/W2            &  0.312  &  0.463  &  0.526  &         &         &  0.50   &  0.33   \\
    \emph{WISE}/W3            &  0.269  &  0.537  &  0.591  &         &         &         &  0.87   \\
    \emph{WISE}/W4            &  0.370  &  0.364  &  0.438  &         &         &         &         \\
    \emph{AKARI}/S9W          &  0.273  &  0.530  &  0.585  &         &         &         &         \\
    \emph{Spitzer}/[3.6]    &  0.260  &  0.553  &  0.605  &  0.63   &  0.56   &         &         \\
    \emph{Spitzer}/[4.5]    &  0.313  &  0.461  &  0.524  &  0.57   &  0.43   &         &         \\
    \emph{Spitzer}/[5.8]    &  0.355  &  0.389  &  0.460  &  0.49   &  0.43   &         &         \\
    \emph{Spitzer}/[8.0]    &  0.334  &  0.426  &  0.492  &  0.55   &  0.43   &         &         \\
    \emph{Spitzer}/[24] &  0.428  &  0.264  &  0.349  &         &         &         &         \\
\tableline
\end{tabular}
\tablenotetext{\ast}{References: (1) GJL09; (2)\citet{Indebetouw05}; (3)\citet{Yuan2013}; (4)\citet{Davenport2014}.}
\label{EAresult}
\end{center}
\end{table}
\end{landscape}

\begin{landscape}
\begin{table}[htbp]
  \centering
  \caption{Results of linear fitting (LF), Bootstrap Re-sampling (BR) and Monte Carlo (MC) simulation}
    \begin{tabular}{ccccccccccc}
    \hline
    \hline
            & W1 & [3.6] & [4.5] & W2 & [5.8] & [8.0] & S9W   & W3  & W4  & [24] \\
    \hline
    $E_{\Ks\lambda}/E_{\rm J\Ks}$ & 0.238  & 0.260  & 0.313  & 0.312  & 0.355  & 0.334  & 0.273  & 0.269  & 0.370  & 0.428  \\
    $\sigma$ (LF) & 2.260E-04 & 0.001  & 0.002  & 2.560E-04 & 0.001  & 0.001  & 0.009  & 0.001  & 0.004  & 0.016  \\
    Mean (BR) & 0.238  & 0.260  & 0.313  & 0.312  & 0.355  & 0.334  & 0.274  & 0.269  & 0.371  & 0.428  \\
    $\sigma$ (BR) & 5.015E-04 & 0.001  & 0.001  & 5.956E-04 & 0.001  & 0.001  & 0.020  & 0.002  & 0.012  & 0.020  \\
    Mean (MC) & 0.236  & 0.259  & 0.312  & 0.310  & 0.354  & 0.332  & 0.268  & 0.265  & 0.360  & 0.426  \\
    $\sigma$ (MC) & 4.592E-04 & 0.001  & 0.001  & 4.518E-04 & 0.001  & 0.001  & 0.015  & 0.001  & 0.010  & 0.007  \\
          &       &       &       &       &       &       &       &       &       &  \\
    Intercept & -0.013  & -0.012  & -0.009  & -0.017  & -0.014  & -0.001  & -0.013  & -0.016  & -0.036  & -0.017  \\
    $\sigma$ (LF)  & 3.660E-04 & 0.001  & 0.001  & 4.170E-04 & 0.001  & 0.001  & 0.021  & 0.001  & 0.010  & 0.014  \\
    Mean (BR) & -0.013  & -0.012  & -0.009  & -0.017  & -0.014  & -0.001  & -0.013  & -0.016  & -0.036  & -0.017  \\
    $\sigma$ (BR) & 1.596E-04 & 0.001  & 0.002  & 1.808E-04 & 0.001  & 0.001  & 0.006  & 3.873E-04 & 0.002  & 0.018  \\
    Mean (MC) & -0.012  & -0.011  & -0.007  & -0.031  & -0.012  & 0.001  & -0.012  & -0.015  & -0.034  & -0.015  \\
    $\sigma$ (MC) & 1.889E-04 & 0.001  & 0.001  & 1.897E-04 & 0.001  & 0.001  & 0.004  & 3.619E-04 & 0.003  & 0.008  \\
    \hline
    \end{tabular}%
  \label{ST_EE}%
\end{table}%
\end{landscape}
\clearpage
\begin{landscape}
\begin{table}
\begin{center}
\caption{\label{binEresult} The $E_{\Ks\lambda}/E_{J\Ks}$ Results within various range of $E_{J\Ks}$ }
\vspace{0.05in}
\begin{tabular}{m{66pt}<{\centering}|ccc|ccc|ccc|ccc}
\tableline \tableline
     & \multicolumn{3}{c|}{all data} & \multicolumn{3}{c|}{$E_{J\Ks}<0.86$} & \multicolumn{3}{c|}{$0.86\leq E_{J\Ks}<1.72$} &  \multicolumn{3}{c}{$E_{J\Ks}\geq 1.72$} \\
     Band & $\frac{E_{\Ks\lambda}}{E_{J\Ks}}$ & sigma & sources & $\frac{E_{\Ks\lambda}}{E_{J\Ks}}$ & sigma & sources & $\frac{E_{\Ks\lambda}}{E_{J\Ks}}$ & sigma & sources & $\frac{E_{\Ks\lambda}}{E_{J\Ks}}$ & sigma & sources \\
\tableline
    \emph{WISE}/W1          &  0.238  & 0.028 & 58143 &  0.207  & 0.029 & 55164 &  0.228  & 0.039 & 2577 &  0.231   & 0.043 & 402 \\
    \emph{WISE}/W2          &  0.312  & 0.032 & 58292 &  0.270  & 0.033 & 55412 &  0.296  & 0.046 & 2476 &  0.302   & 0.051 & 404 \\
    \emph{WISE}/W3          &  0.269  & 0.060 & 37353 &  0.227  & 0.060 & 36416 &  0.251  & 0.087 &  840 &  0.261   & 0.089 &  97 \\
    \emph{WISE}/W4          &  0.370  & 0.076 &  1481 &  0.258  & 0.080 &  1458 &  0.350  & 0.108 &   22 &   NaN    &  NaN  &   1 \\
    \emph{AKARI}/S9W        &  0.273  & 0.156 &  1007 &  0.243  & 0.157 &   986 &  0.264  & 0.140 &   18 &  0.246   & 0.103 &   3 \\
    \emph{Spitzer}/[3.6]    &  0.260  & 0.054 &  5327 &  0.254  & 0.054 &  2982 &  0.260  & 0.054 & 1797 &  0.258   & 0.053 & 548 \\
    \emph{Spitzer}/[4.5]    &  0.313  & 0.061 &  5466 &  0.320  & 0.061 &  3072 &  0.316  & 0.062 & 1830 &  0.314   & 0.062 & 564 \\
    \emph{Spitzer}/[5.8]    &  0.355  & 0.050 &  5373 &  0.355  & 0.049 &  3010 &  0.356  & 0.051 & 1804 &  0.355   & 0.053 & 559 \\
    \emph{Spitzer}/[8.0]    &  0.334  & 0.052 &  5391 &  0.352  & 0.049 &  3057 &  0.346  & 0.054 & 1790 &  0.338   & 0.063 & 544 \\
    \emph{Spitzer}/[24]     &  0.428  & 0.255 &   786 &  0.423  & 0.231 &   469 &  0.416  & 0.272 &  229 &  0.415   & 0.330 &  88 \\
\tableline
\end{tabular}
\end{center}
\end{table}
\end{landscape}

\begin{landscape}
\begin{table}
\begin{center}
\caption{\label{binAresult} The $A_{\lambda}/A_{\Ks}$ Results within various range of $A_{\Ks}$}
\vspace{0.02in}
\begin{tabular}{lccccccccc}
\tableline \tableline
                       & \multicolumn{3}{c}{This Work} & \multicolumn{4}{c}{\cite{Chapman2009}} & \multicolumn{2}{c}{\cite{McClure2009}} \\
  $A_{\Ks}$ Range     & $<0.5$ & $[0.5,1]$ & $>1$ & $<0.5$ & $[0.5,1]$ & $[1,2]$ & $>2$ & $[0.3,1]$ & $[1,7]$ \\
\tableline
    \emph{WISE}/W1        & 0.644 & 0.608 & 0.603  &        &        &        &        &        &         \\
    \emph{WISE}/W2        & 0.536 & 0.491 & 0.481  &        &        &        &        &        &         \\
    \emph{WISE}/W3        & 0.610 & 0.568 & 0.551  &        &        &        &        &  0.415 &  0.534  \\
    \emph{WISE}/W4        & 0.556 & 0.398 &  NaN   &        &        &        &        &  0.260 &  0.429  \\
    \emph{AKARI}/S9W      & 0.582 & 0.546 & 0.577  &        &        &        &        &  0.428 &  0.493  \\
    \emph{Spitzer}/[3.6]  & 0.563 & 0.553 & 0.556  &  0.41  &  0.49  &  0.60  &  0.64  &  0.630 &  0.630  \\
    \emph{Spitzer}/[4.5]  & 0.450 & 0.456 & 0.460  &  0.26  &  0.35  &  0.46  &  0.53  &  0.530 &  0.530  \\
    \emph{Spitzer}/[5.8]  & 0.389 & 0.388 & 0.389  &  0.28  &  0.35  &  0.44  &  0.46  &  0.381 &  0.477  \\
    \emph{Spitzer}/[8.0]  & 0.395 & 0.405 & 0.419  &  0.21  &  0.35  &  0.43  &  0.45  &  0.363 &  0.450  \\
    \emph{Spitzer}/[24]   & 0.272 & 0.284 & 0.286  &  1.08  &  0.75  &  0.61  &  0.34  &  0.239 &  0.404  \\
\tableline
\end{tabular}
\end{center}
\end{table}
\end{landscape}

\begin{figure}
\centering
\includegraphics[scale=0.6]{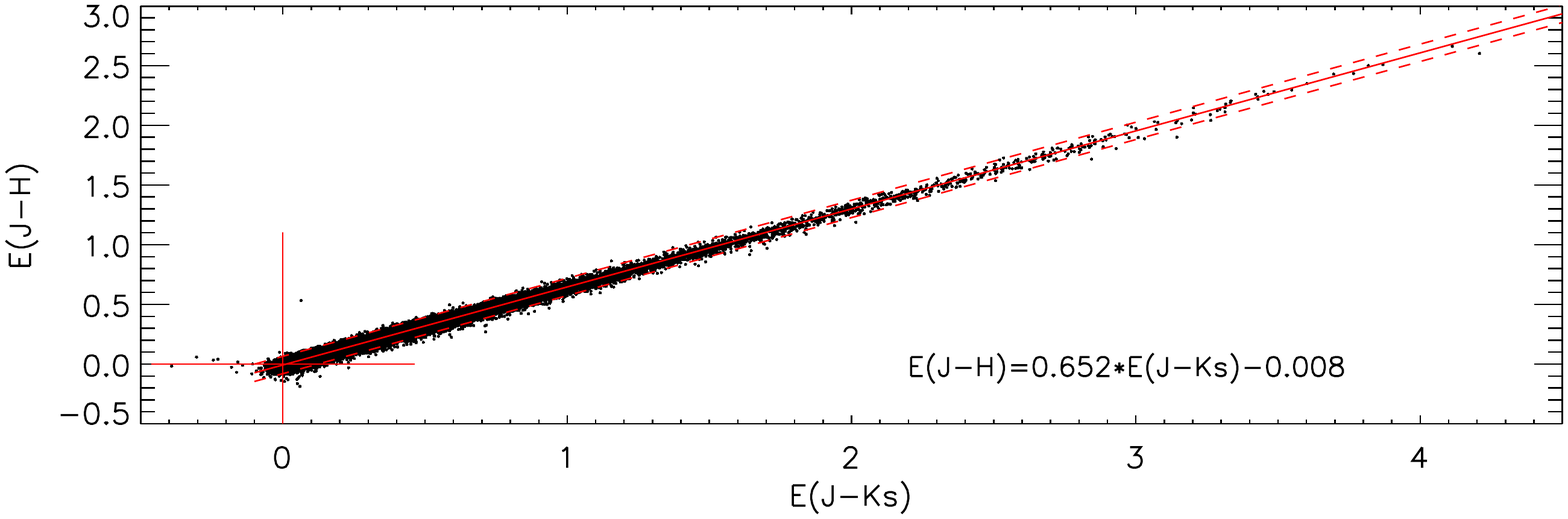}
\caption{
         Linear fitting between the color-excess ratio $E_{\rm JH}$ and $E_{\rm J\Ks}$. The fitting line (red solid line) and equation are displayed, together with the 3$\sigma$ range (red dash lines). Also shown is the point (0.0, 0.0) for zero-reddening.
\label{fig4.2}}
\end{figure}

\begin{figure}
\centering
\includegraphics[scale=0.6]{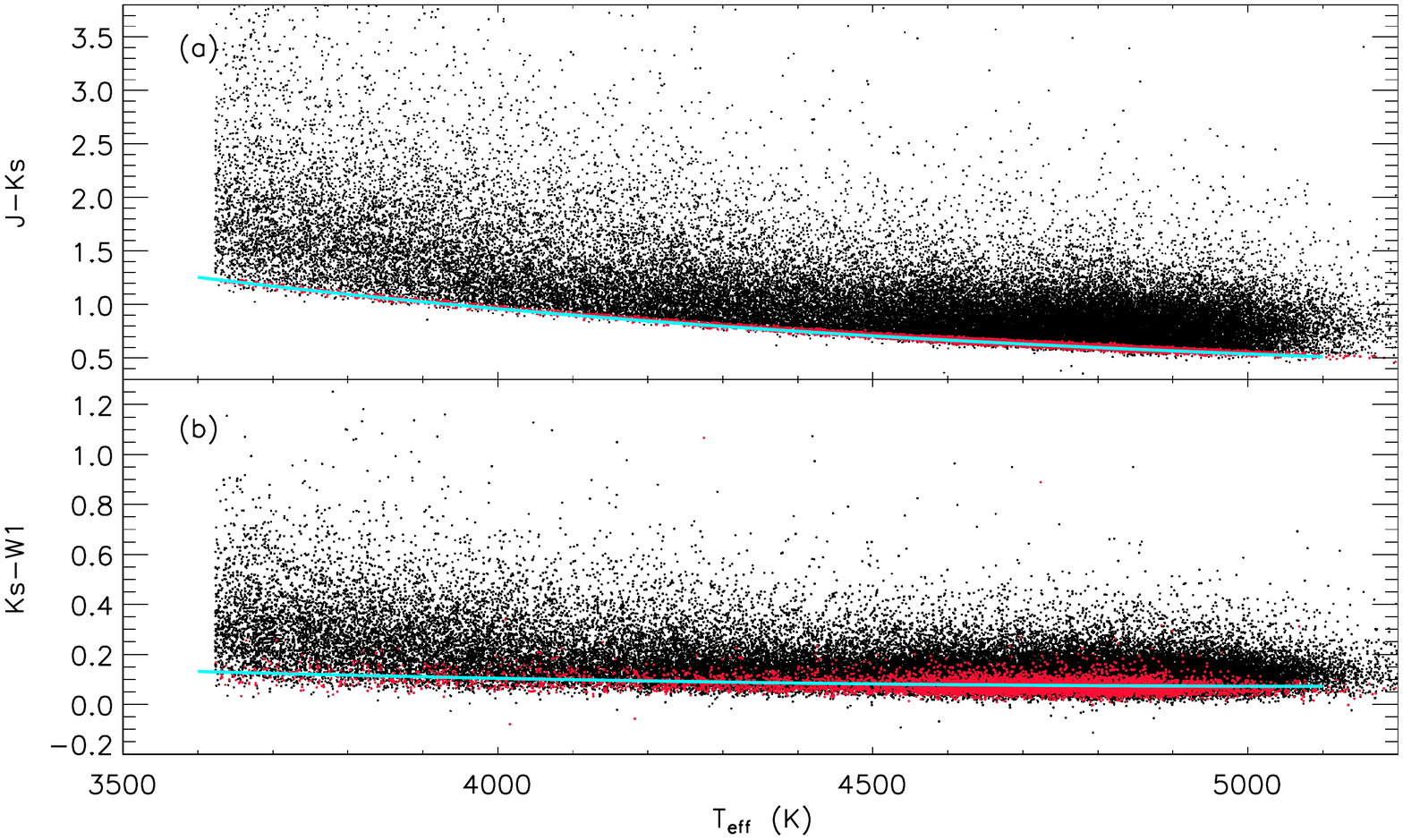}
\caption{
          Same as Figure~\ref{fig4.1}, but for the color index $C_{\rm{KsW1}}$ and for the sources cross-identified between APOGEE and the \emph{WISE}/W1 band. The following figures (Figs 5-13)  are for the sources cross-identified in the corresponding waveband.
\label{fig41}}
\end{figure}

\begin{figure}
\centering
\includegraphics[scale=0.6]{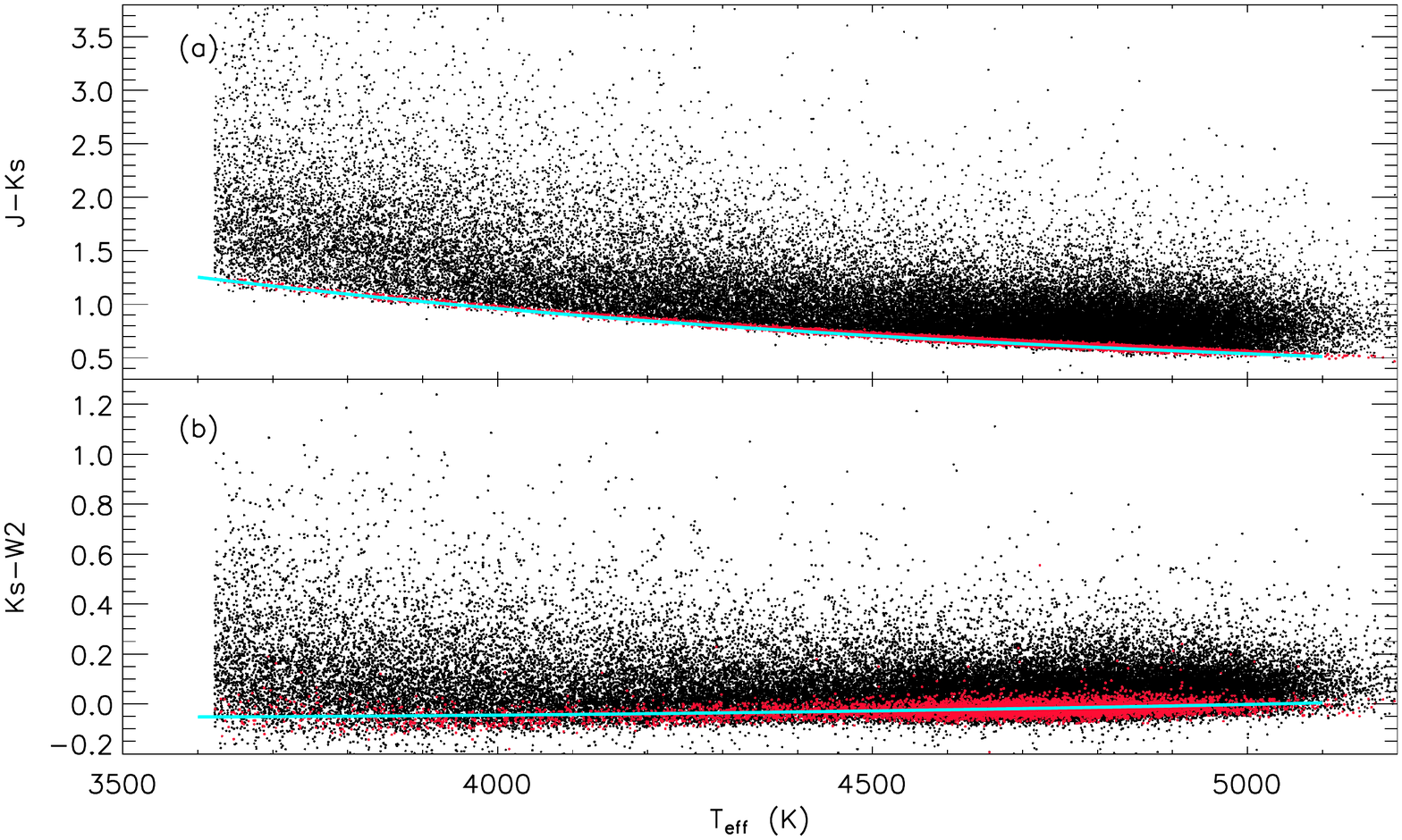}
\caption{
          Same as Figure~\ref{fig4.1}, but for the color index $C_{\rm{KsW2}}$.
\label{fig42}}
\end{figure}

\begin{figure}
\centering
\includegraphics[scale=0.6]{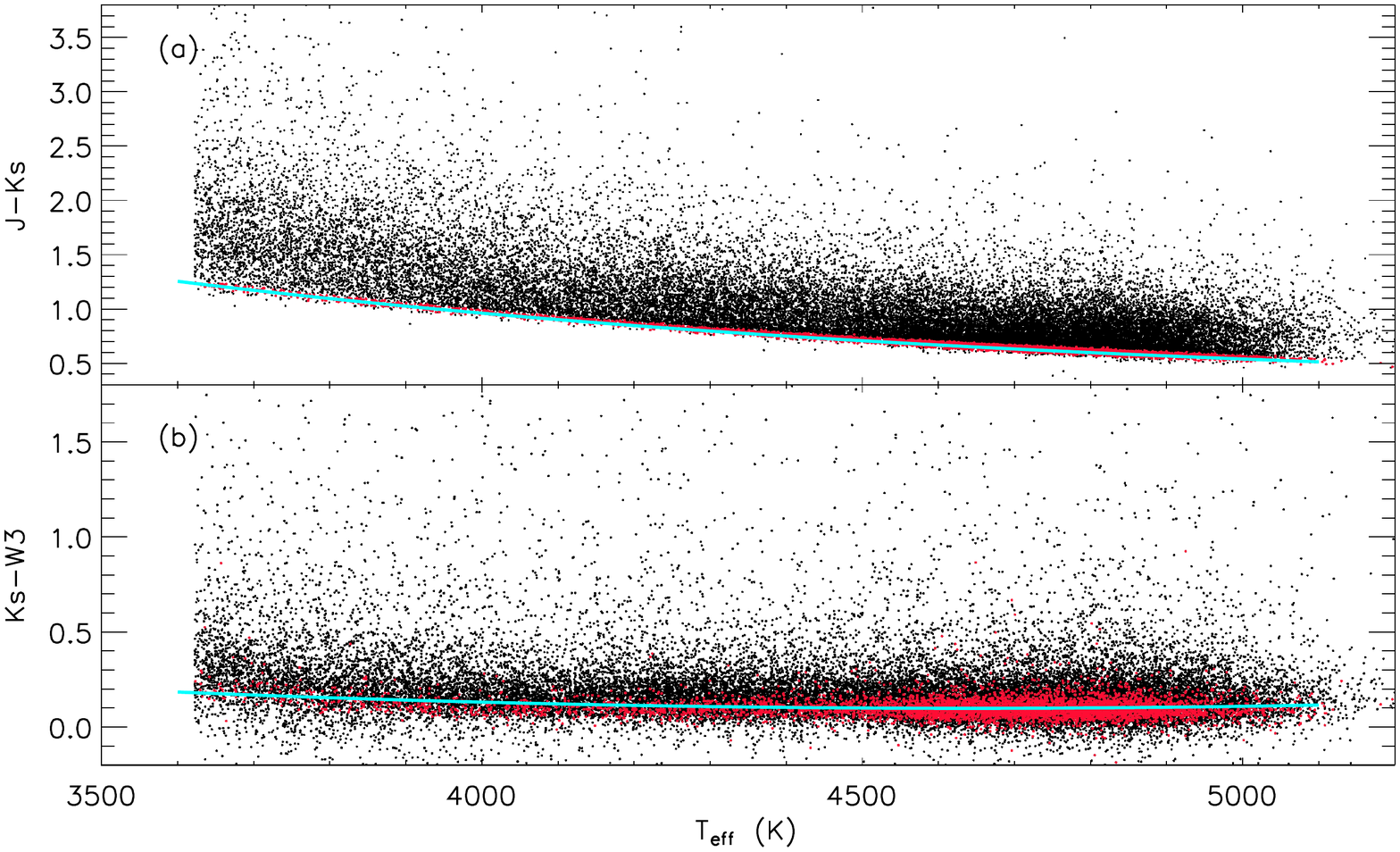}
\caption{Same as Figure~\ref{fig4.1}, but for the color index $C_{\rm{KsW3}}$.
\label{fig43}}
\end{figure}

\begin{figure}
\centering
\includegraphics[scale=0.6]{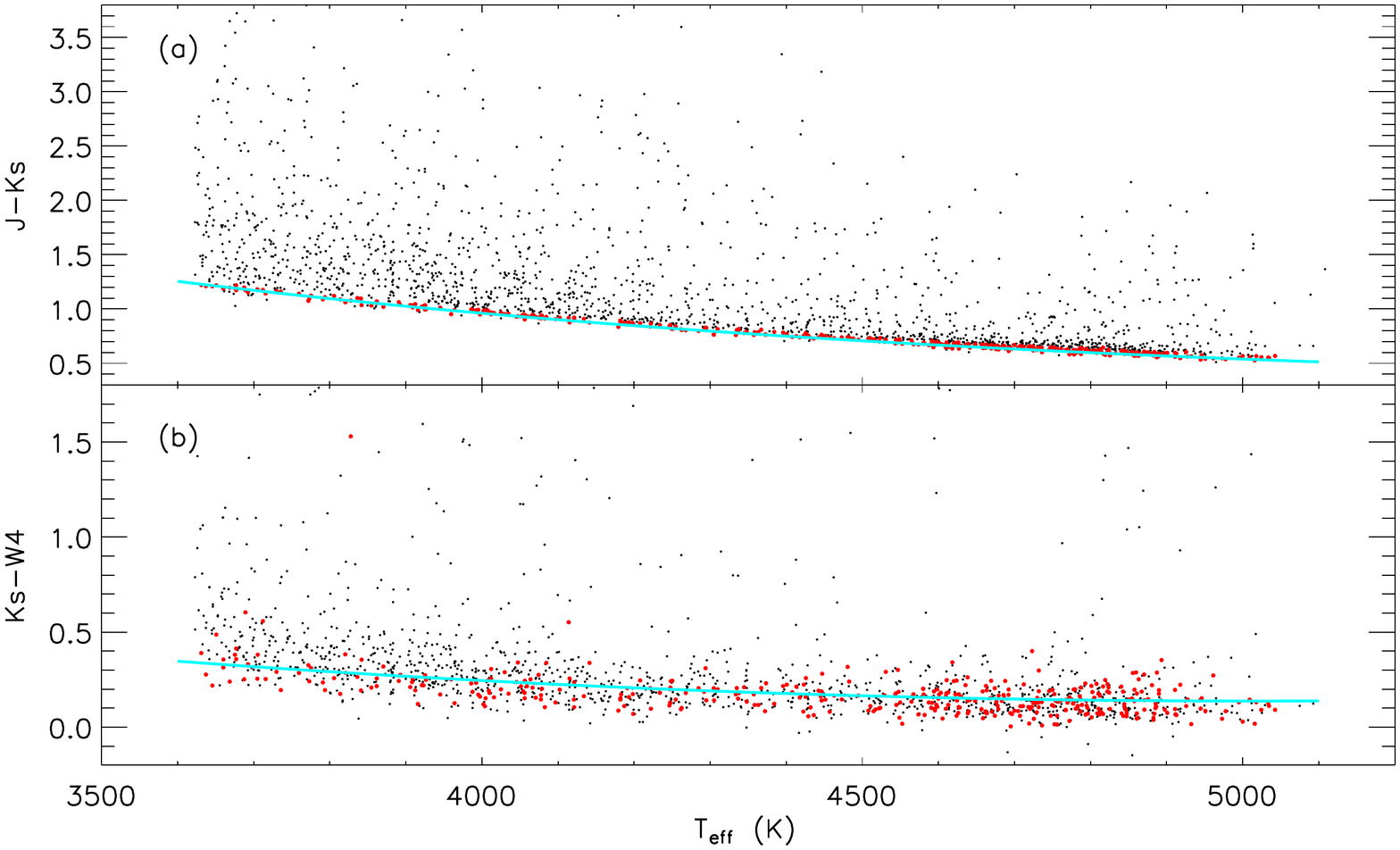}
\caption{Same as Figure~\ref{fig4.1}, but for the color index $C_{\rm{KsW4}}$.
\label{fig44}}
\end{figure}

\begin{figure}
\centering
\includegraphics[scale=0.6]{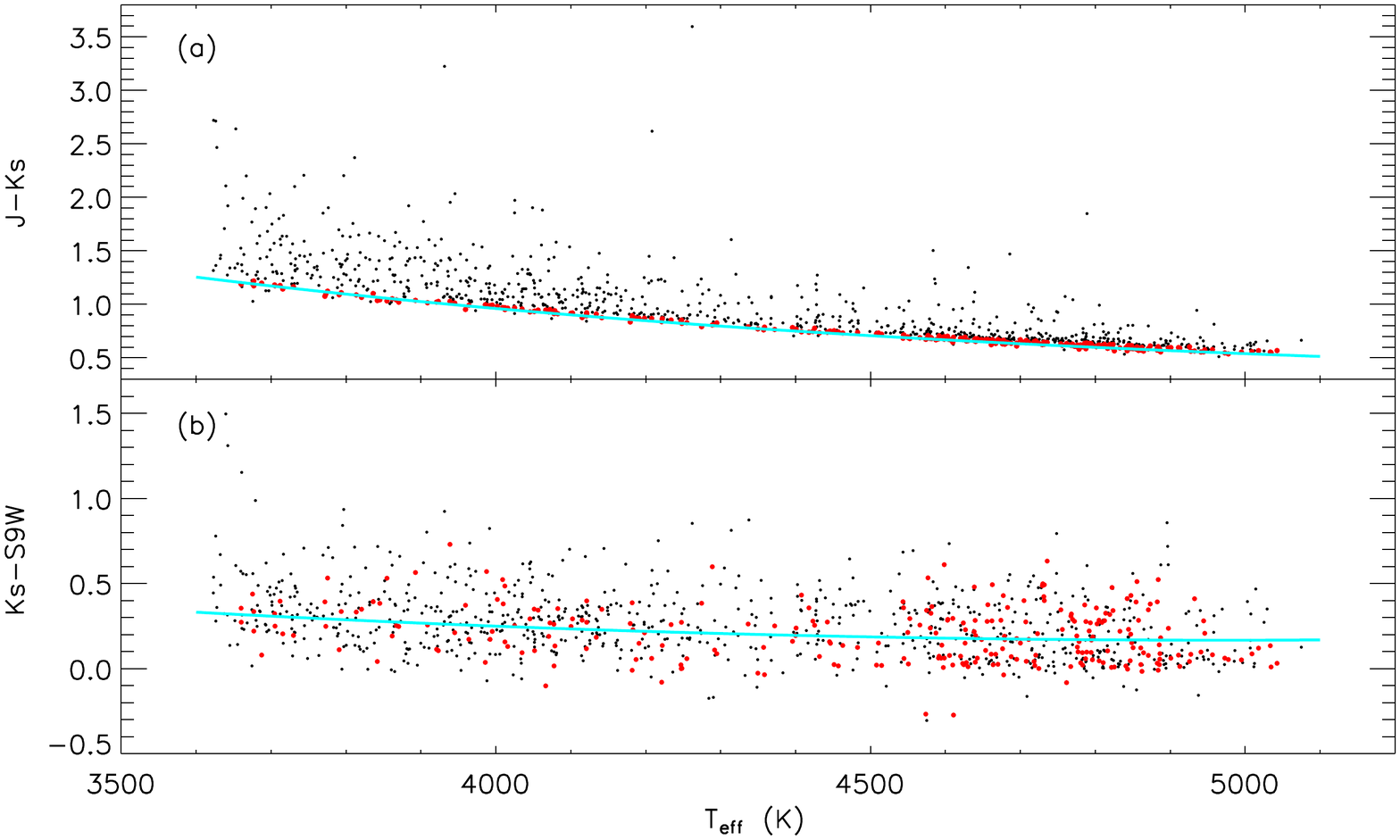}
\caption{Same as Figure~\ref{fig4.1}, but for the color index $C_{\rm{KsS9W}}$.
\label{fig3}}
\end{figure}

\begin{figure}
\centering
\includegraphics[scale=0.6]{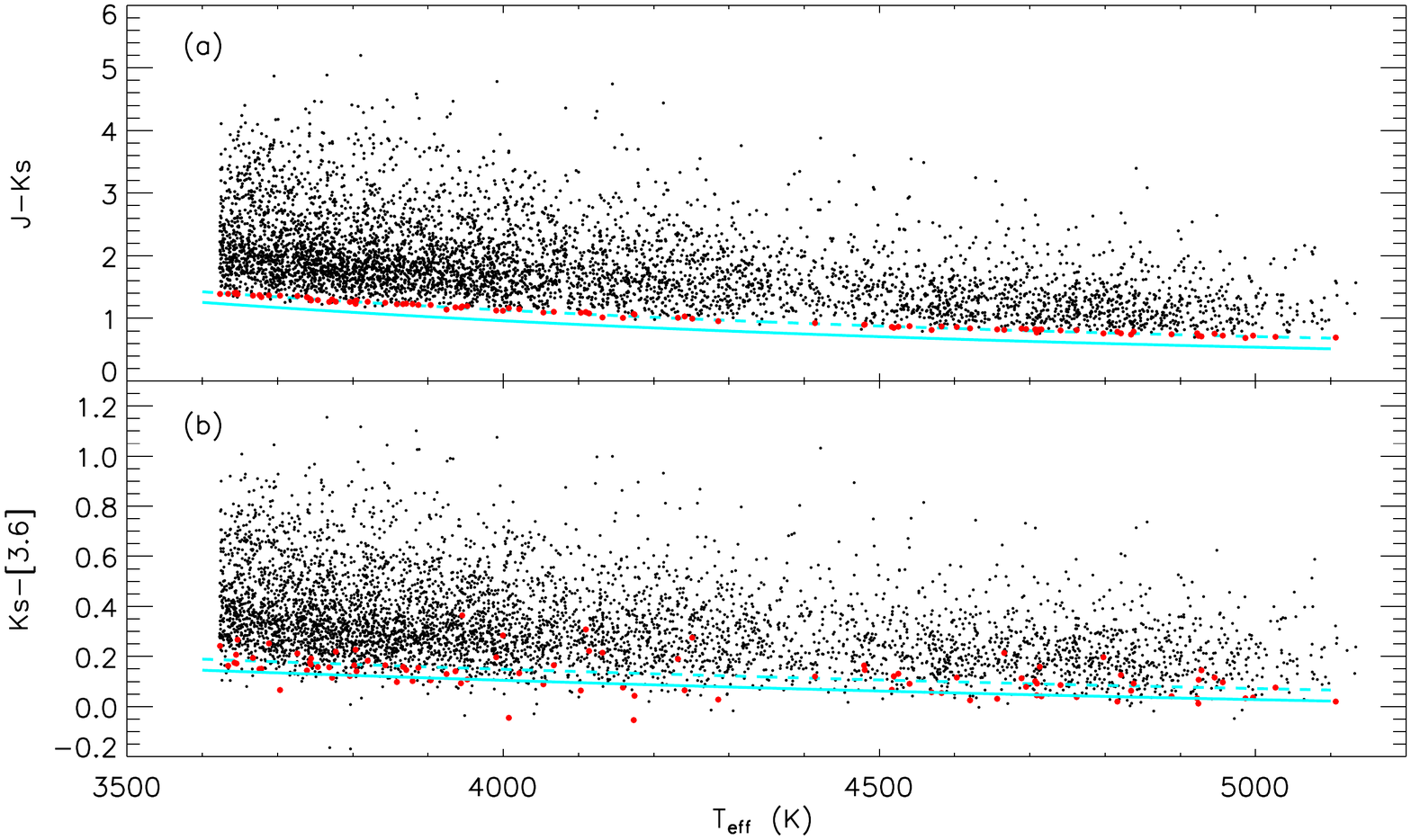}
\caption{Same as Figure~\ref{fig4.1}, but for the color index $C_{\rm{Ks[3.6]}}$.
\label{fig31}}
\end{figure}

\clearpage

\begin{figure}
\centering
\includegraphics[scale=0.6]{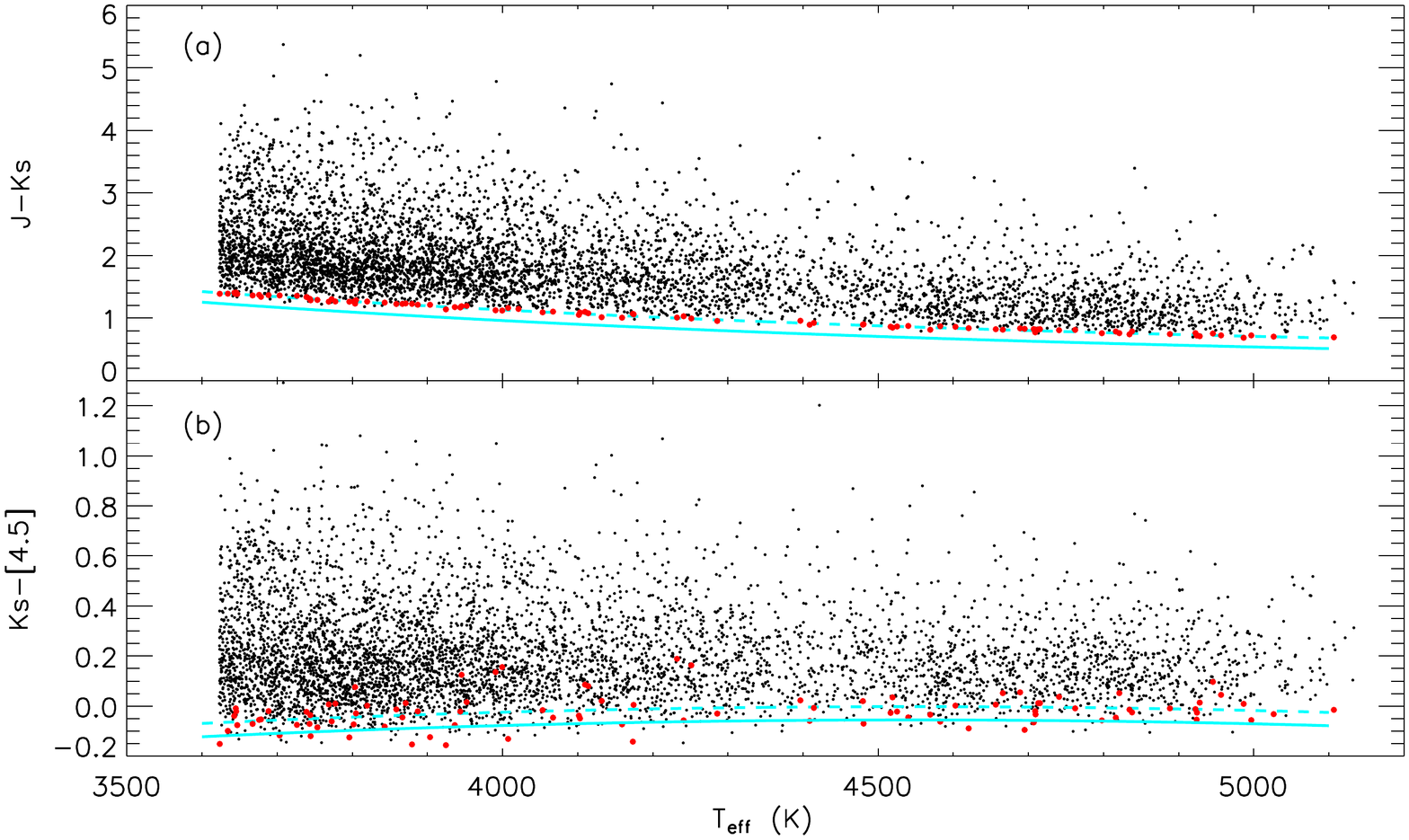}
\caption{Same as Figure~\ref{fig4.1}, but for the color index $C_{\rm{Ks[4.5]}}$
\label{fig32}}
\end{figure}

\begin{figure}[htbp]
\centering
\includegraphics[scale=0.6]{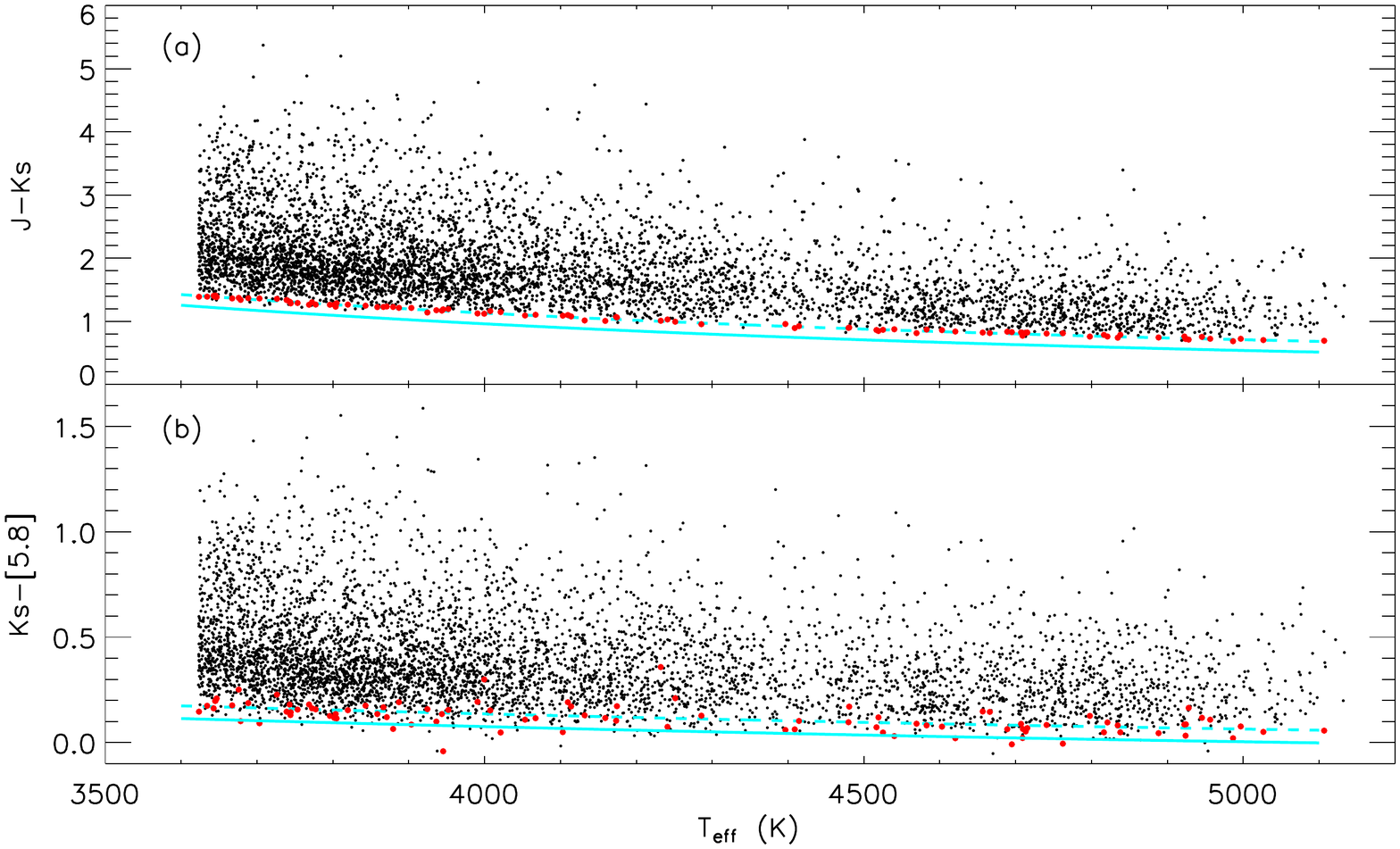}
\caption{Same as Figure~\ref{fig4.1}, but for the color index $C_{\rm{Ks[5.8]}}$
\label{fig33}}
\end{figure}

\begin{figure}
\centering
\includegraphics[scale=0.6]{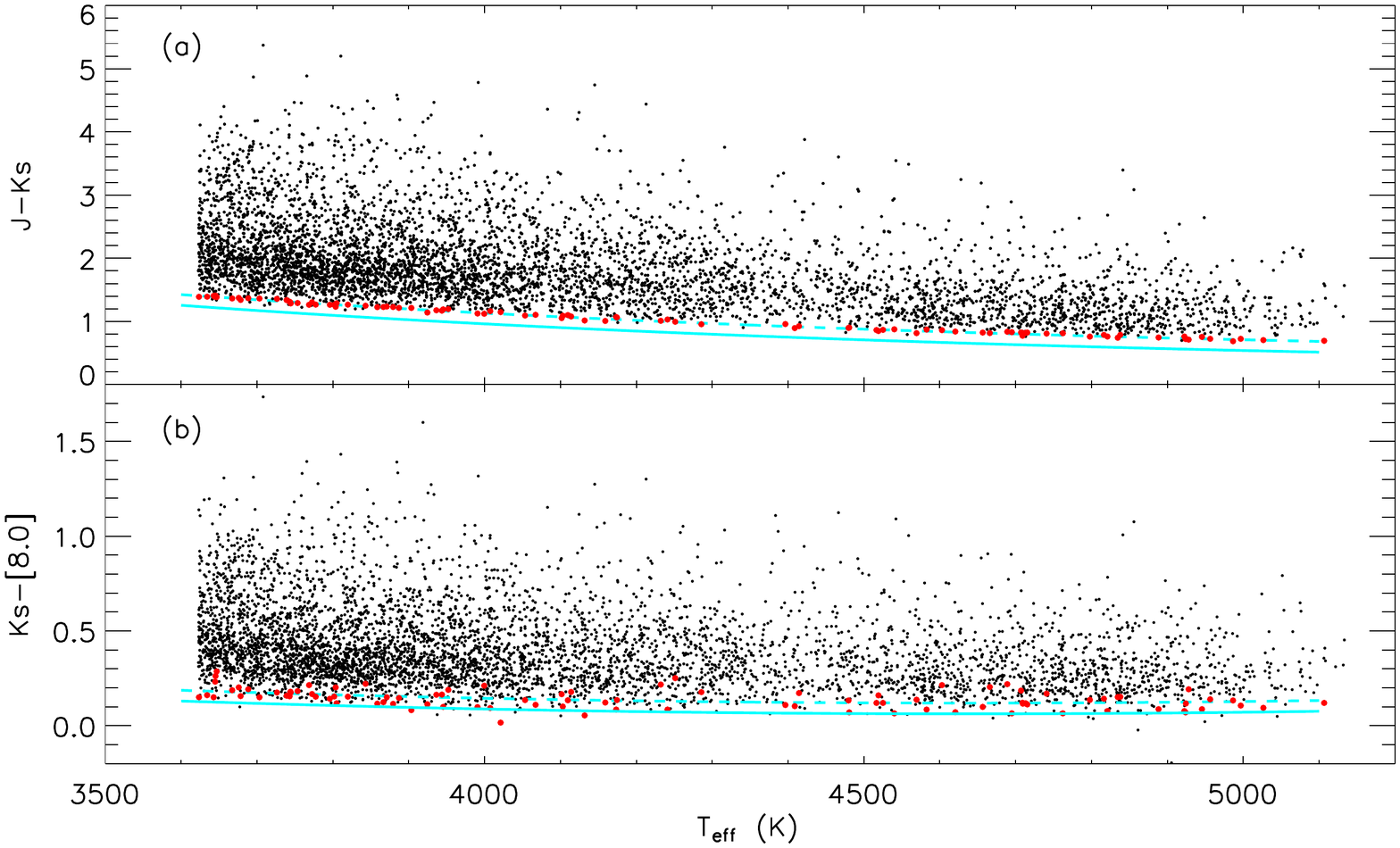}
\caption{Same as Figure~\ref{fig4.1}, but for the color index $C_{\rm{Ks[8.0]}}$
\label{fig34}}
\end{figure}

\begin{figure}
\centering
\includegraphics[scale=0.6]{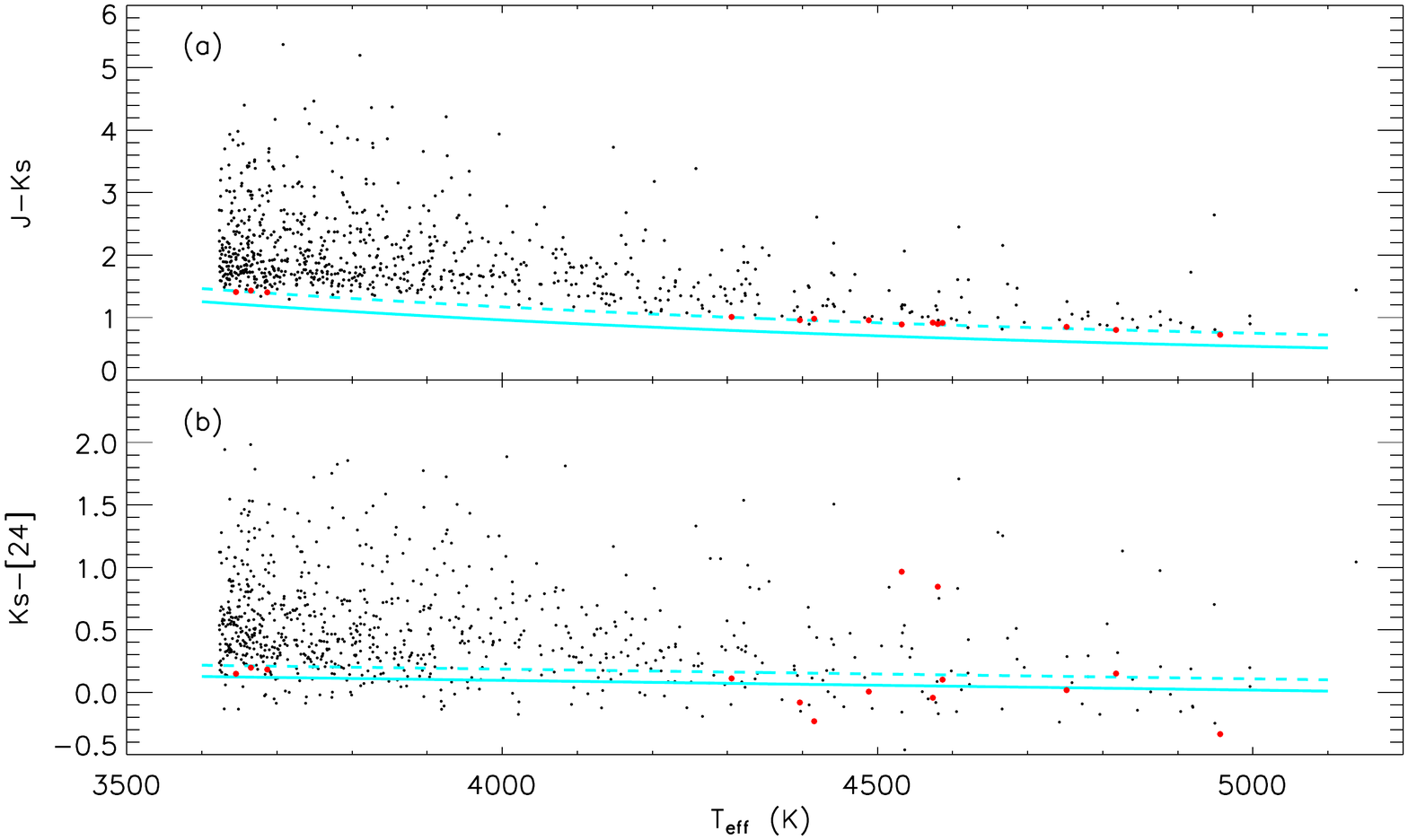}
\caption{Same as Figure~\ref{fig4.1}, but for the color index $C_{\rm{Ks[24]}}$
\label{fig35}}
\end{figure}

\begin{figure}
\centering
\includegraphics[scale=0.6]{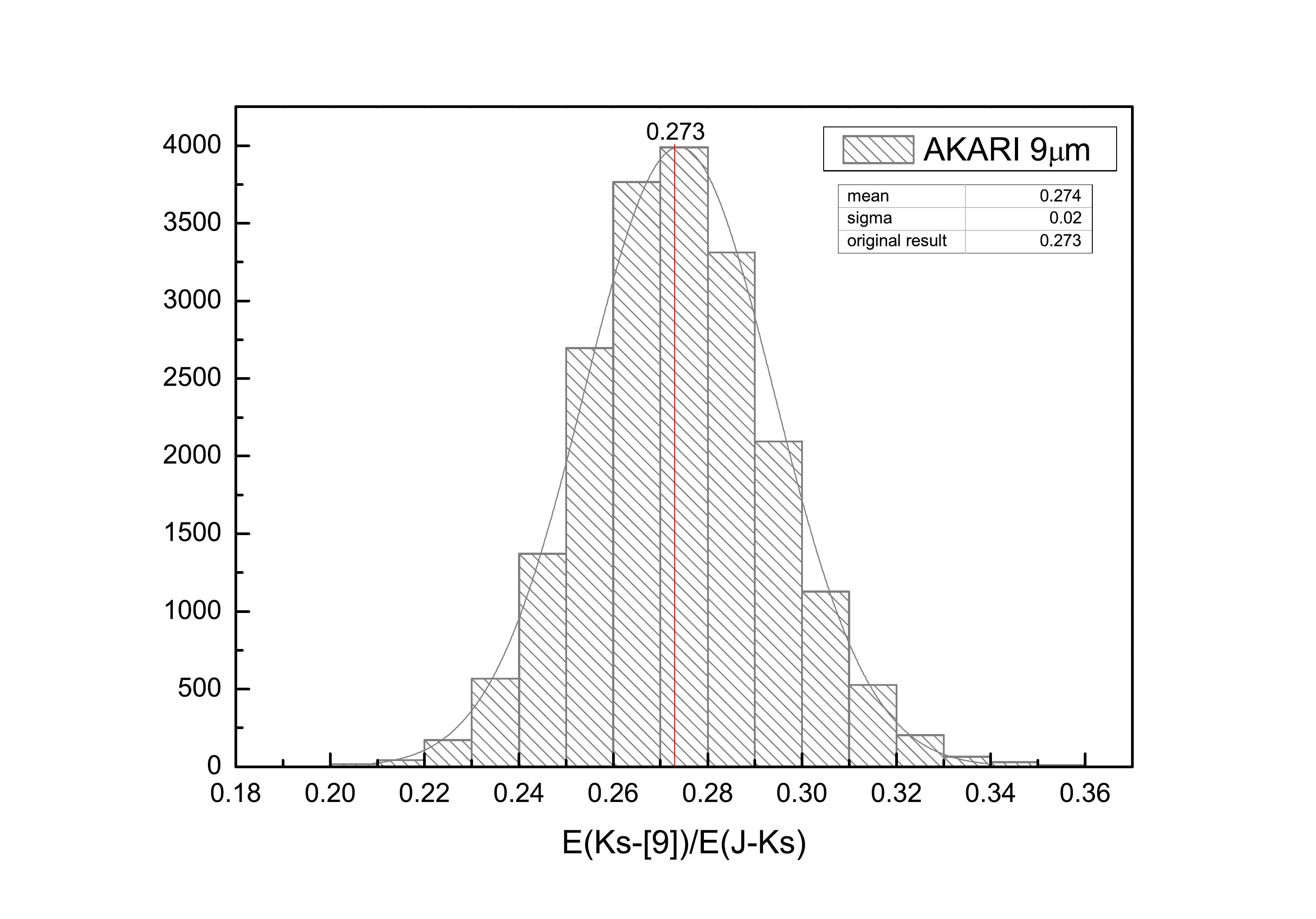}
\caption{
          The slope distribution of a 20000-time Bootstrap resampling test of $E_{\rm \Ks S9W}/E_{\rm J\Ks}$. The red line indicates the linear fitting result of $E_{\rm \Ks S9W}/E_{\rm J\Ks}$ as given in Table \ref{EAresult}.
\label{BootstrapE9}}
\end{figure}

\begin{figure}
\centering
\includegraphics[scale=0.7]{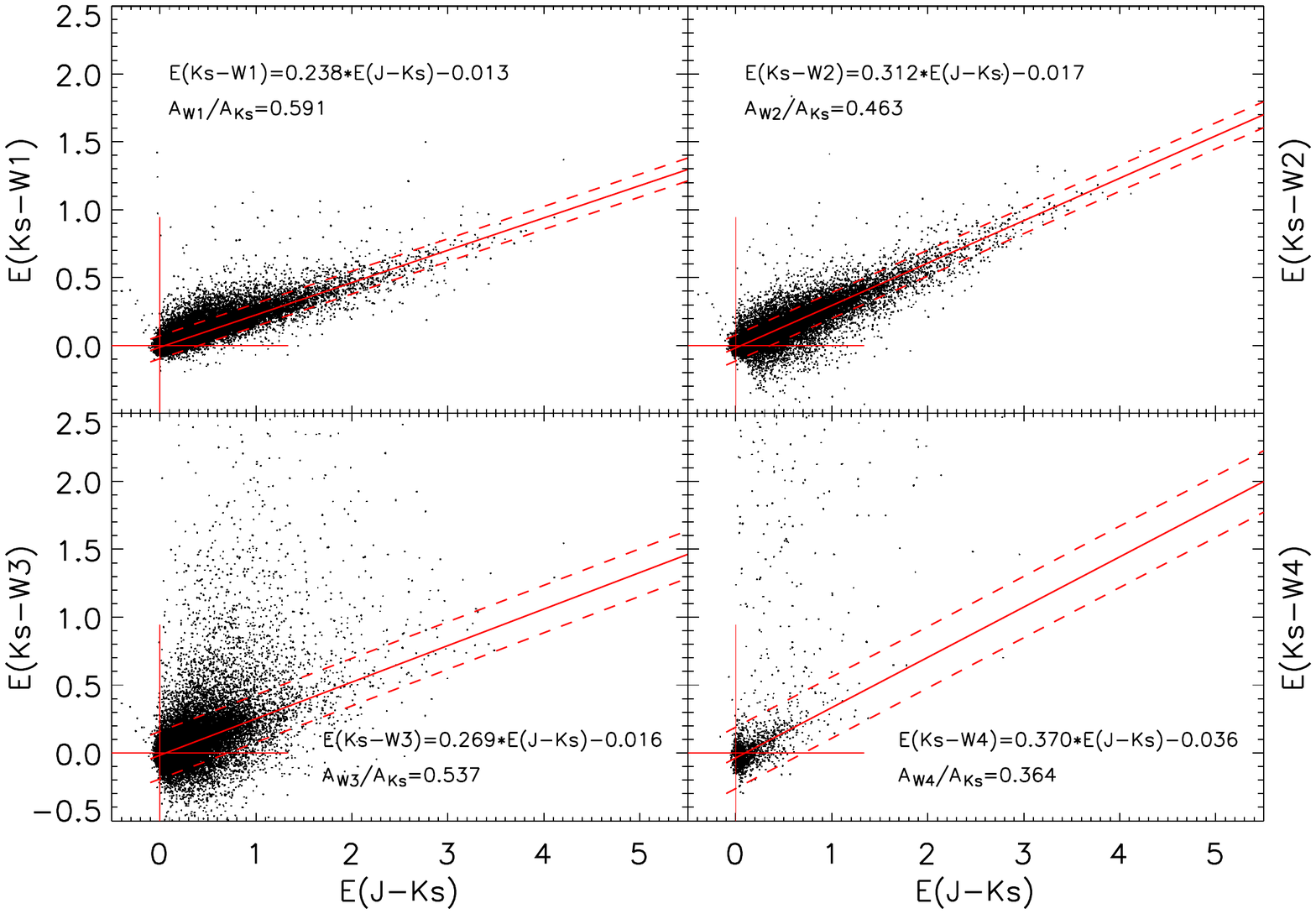}
\caption{
         The color-excess vs. color-excess diagram. The symbol conventions are the same as in Figure~\ref{fig4.2}. For the four \emph{WISE} bands. We adopt $A_{\rm J}/A_{\rm \Ks}=2.72$ (see \S\ref{sec_nir_ext}).
\label{JK_KW1234}}
\end{figure}

\begin{figure}
\centering
\includegraphics[scale=0.7]{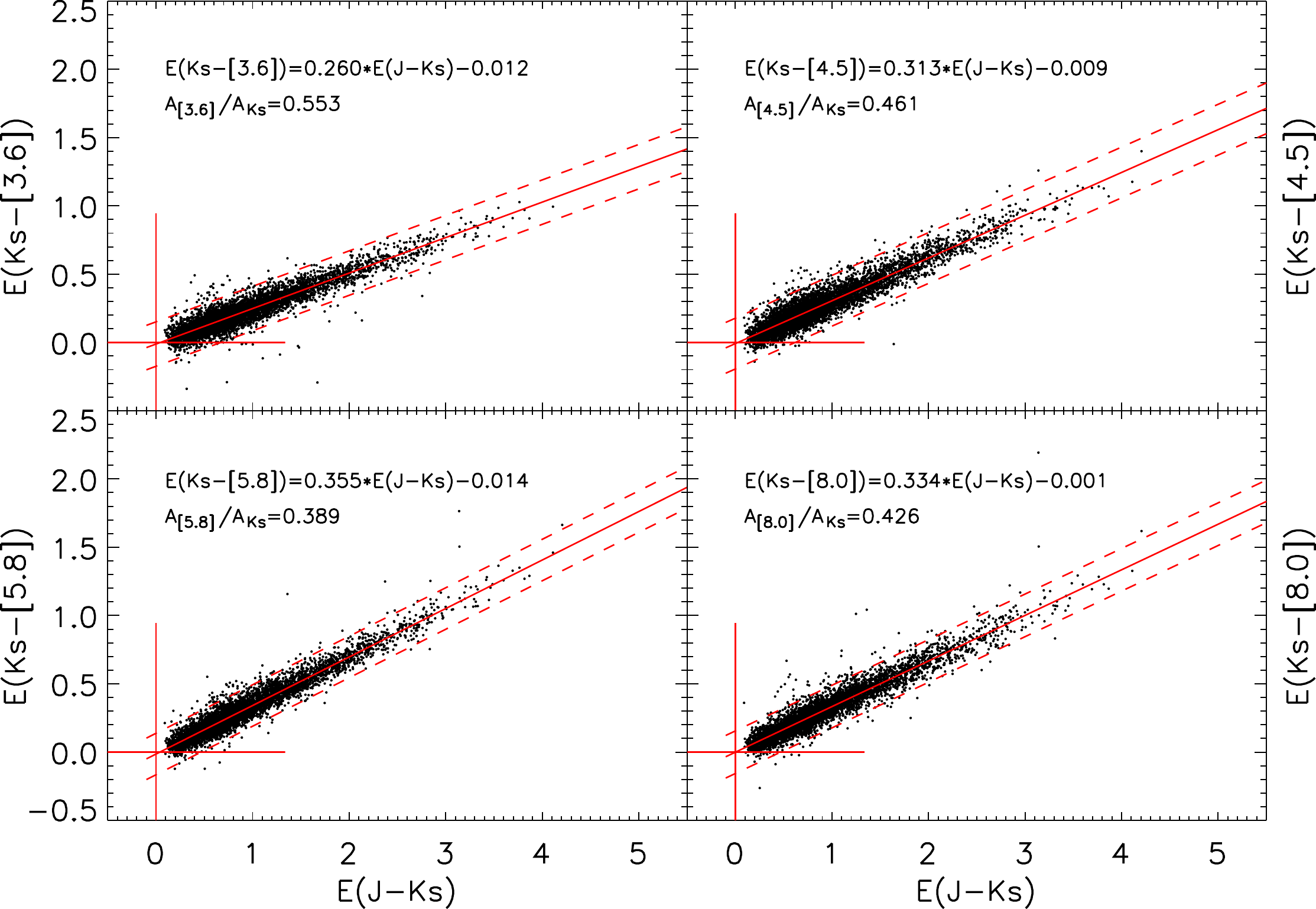}
\caption{Same as Figure~\ref{JK_KW1234}, but for the four \emph{Spitzer}/IRAC bands.
\label{JK_KIRAC}}
\end{figure}

\begin{figure}
\centering
\includegraphics[scale=0.7]{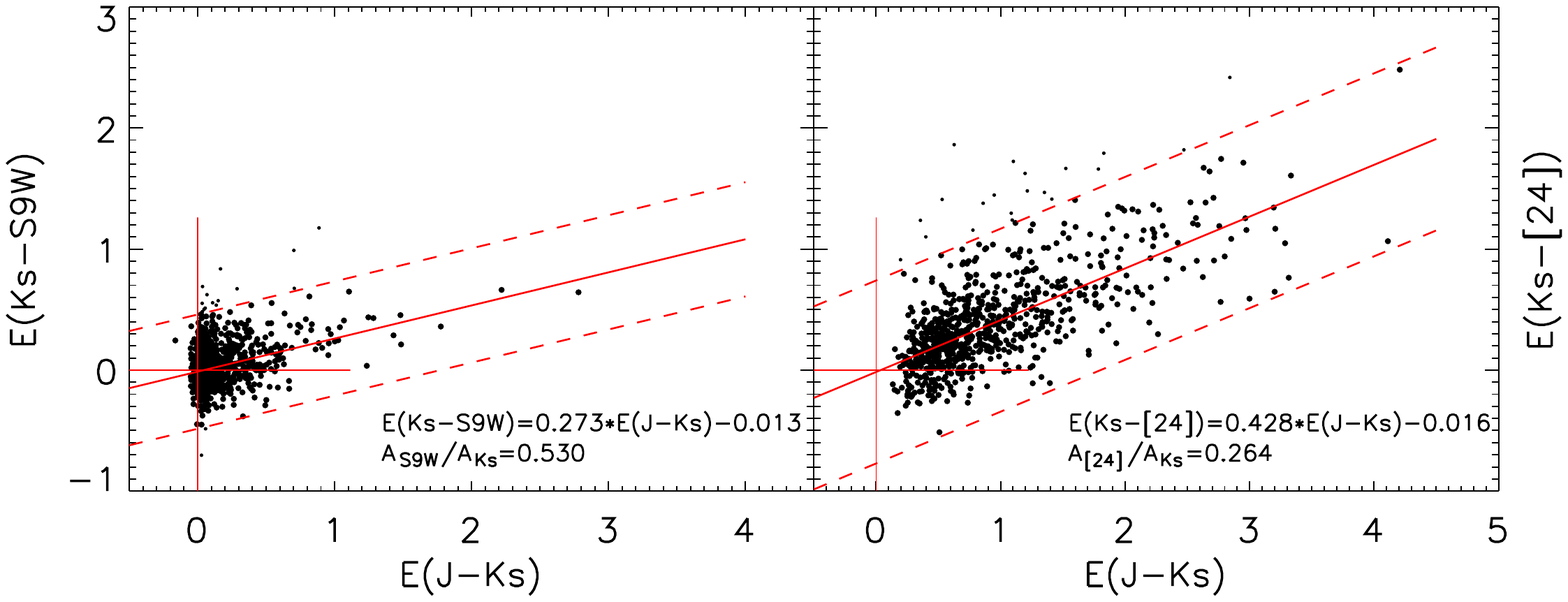}
\caption{Same as Figure~\ref{JK_KW1234}, but for the \emph{AKARI}/S9W and \emph{Spitzer}/MIPS [24] bands.
\label{JK_K9_24}}
\end{figure}

\clearpage

\begin{figure}
\centering
\includegraphics[scale=0.6]{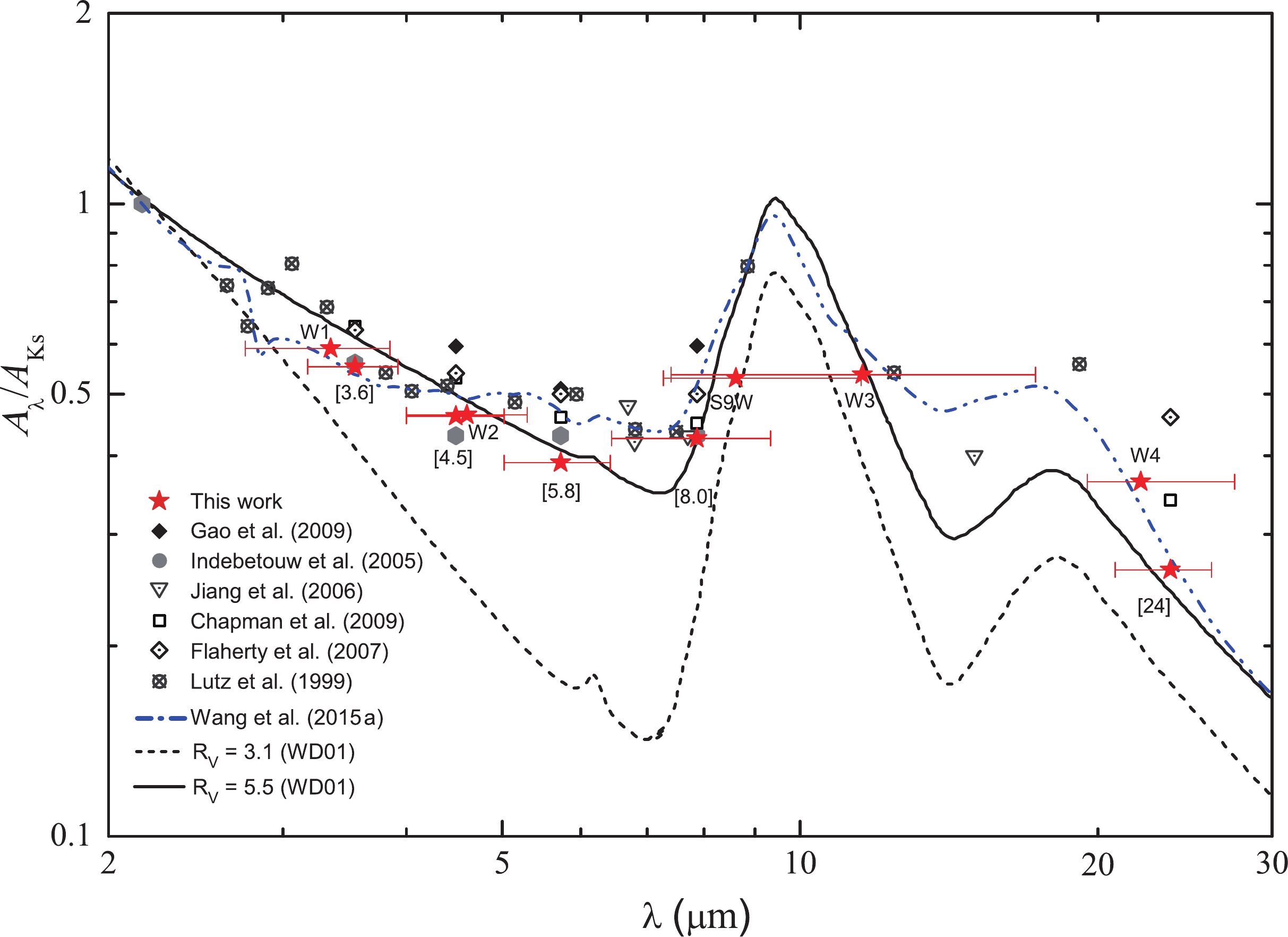}
\caption{
          Comparison of the extinction derived in this work
          (red stars) with the previous determinations
          and model predictions.
          The horizontal bars show the wavelength coverages
           of the photometric bands.
\label{extcurve}}
\end{figure}

\begin{figure}
\centering
\includegraphics[scale=0.6]{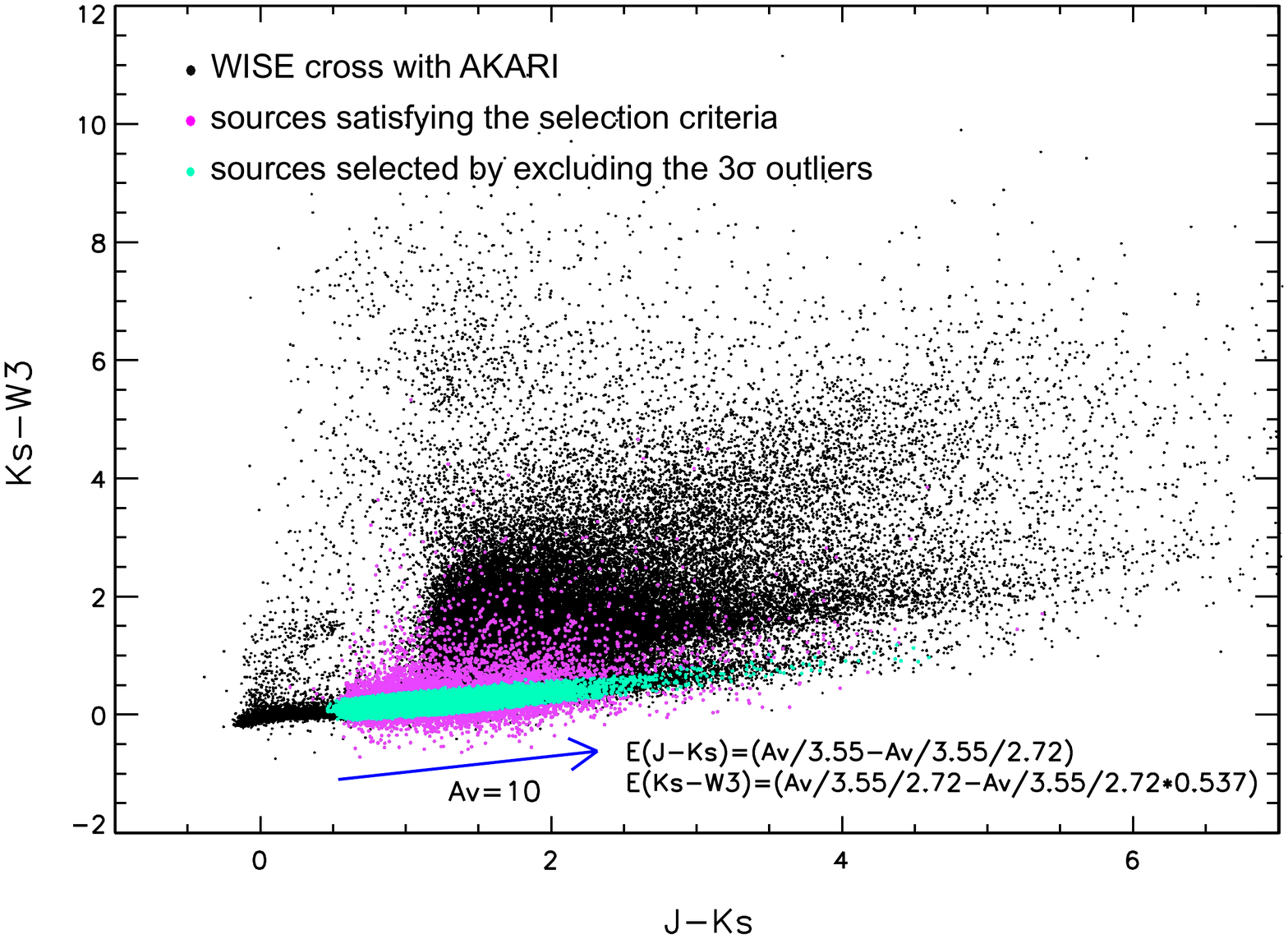}
\caption{
          Color-color diagram of $C_{\rm \Ks W3}$ and $C_{\rm J\Ks}$. Black dots are the sources cross-identified between \emph{AKARI}/S9W and \emph{WISE}/W3, purple dots are the stars meet the selection criteria in \S\ref{sec_quality_control}, and the blue dots are the stars which are finally selected to determine the color-excess ratio after excluding the 3$\sigma$ outliers.
\label{JK_KW3_liu_xue}}
\end{figure}

\begin{figure}
\centering
\includegraphics[scale=0.6]{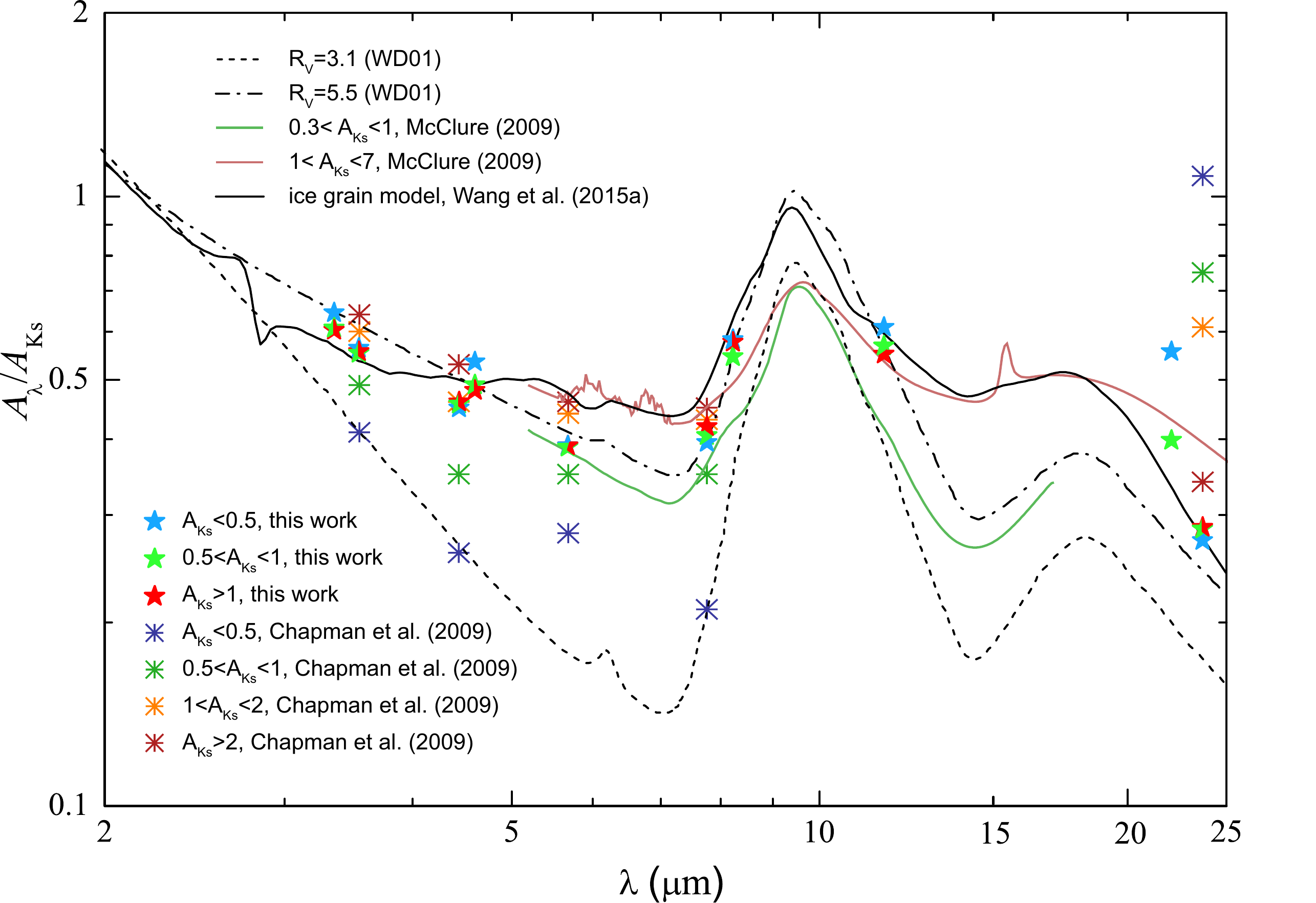}
\caption{
          IR extinction laws in various $\AKs$ ranges.
\label{bin_curve}}
\end{figure}
\end{document}